\documentclass[prd,aps,epsig]{revtex4}
\usepackage{graphicx}
\usepackage{dcolumn}
\usepackage{bm}
\usepackage{feynmp}

\newcommand{\bea}{\begin{eqnarray}}
\newcommand{\eea}{\end{eqnarray}}
\begin{document}
\begin{flushright}
TUHEP-TH-05151
\end{flushright}
\title{QCD Multipole Expansion and Hadronic Transitions \\ in  Heavy Quarkonium Systems}
\author{Yu-Ping Kuang}
\address{Center for High Energy Physics and Department of Physics, Tsinghua University, 
Beijing 100084, P.R. of China}
\begin{abstract}
We review the developments of QCD multipole expansion and its applications to hadronic 
transitions and some radiative decays of heavy quarkonia. Theoretical predictions are 
compared with updated experimental results.
\end{abstract}
\maketitle
\section{Introduction}

Heavy quarkonia are the simplest objects for studying the physics of hadrons due to their 
nonrelativistic nature. Although the spectra of heavy quarkonium systems $c\bar{c}$ and 
$b\bar{b}$ have been successfully explained by certain QCD motivated potential models, some of their 
decays concerning nonperturbative QCD are difficult to deal with. 
Hadronic transitions
\begin{eqnarray}                       
\Phi_I\to\Phi_F+h
\label{HT}
\end{eqnarray}
are of this kind. In (\ref{HT}), $\Phi_I$, $\Phi_F$ and $h$ stand for the initial state quarkonium,
the  final state quarkonium, and the emitted light hadron(s),
respectively. Hadronic transitions are important decay modes of heavy quarkonia. For instance, the 
branching ratio for $\psi^\prime\to J/\psi+\pi+\pi$ is approximately $50\%$. 

In the $c\bar{c}$ and $b\bar{b}$ systems, the typical mass difference
$M_{\Phi_I}-M_{\Phi_F}$ is around a few hundred MeV, so that the
typical momentum of the light hadron(s) $h$ is low. So far as the coupled-channel
effect is not concerned, the light hadron(s) $h$ are converted
from the gluons emitted by the heavy quark $Q$ or antiquark
$\bar{Q}$ in the transition. So the typical momentum of the
emitted gluons is also low, and thus perturbative QCD does not
work in these processes. Certain nonperturbative approaches are
thus needed for studying hadronic transitions. In this article,
we review the theoretical framework and applications of 
a feasible approach, {\it QCD multipole expansion} (QCDME),
which works quite well in predicting hadronic transition rates in the $c\bar{c}$ and $b\bar{b}$
systems. In addition to hadronic transitions, QCDME can also lead to successful results in certain 
radiative decay processes such as $J/\psi\to\gamma\eta$ and $J/\psi\to\gamma\eta'$. 
   
   This paper is organized as follows. In Sec. II, we review the theoretical
framework and the formulation of QCDME. Sec. III deals with applications
of QCDME to various hadronic transition processes in the nonrelativistic 
single-channel approach including hadronic transitions between $S$-wave quarkonia, between $P$-wave 
quarkonia, $\pi\pi$ transition of the $D$-wave quarkonia, and the search for the spin-singlet $P$-wave 
quarkonium $h_c$ through hadronic transition. Sec. IV is on the nonrelativistic coupled-channel
theory of hadronic transitions. In Sec, V, we show how QCDME makes successful
predictions for the radiative decays $J/\psi\to \gamma\eta$ and $J/\psi\to\gamma\eta'$,
etc. A summary is given in Sec. VI.

\section{QCD Multipole Expansion}

Multipole expansion in electrodynamics has been widely used for studying radiation processes
in which the electromagnetic field is radiated from local sources. If the radius $a$ of a local 
source is smaller than the wave length $\lambda$ of the radiated electromagnetic field such that
$a/\lambda\sim ak<1$ ($k$ stands for the momentum of the photon), $ak$ can be a good expansion 
parameter, i.e., we can expand the electromagnetic field in powers of $ak$. This is the well-known 
multipole expansion. In classical electrodynamics, the coefficient of the $(ak)^l$ term in the
multipole expansion contains a factor $\displaystyle\frac{1}{(2l+1)!!}$. 
Hence multipole expansion actually works better than what is expected by simply estimating the size 
of $(ak)^l$.

Due to the nonrelativistic nature of heavy quarkonia, the bound states of a heavy quark 
$Q$ and its antiquark $\bar{Q}$ can be calculated by solving the Schr\"odinger equation with a 
given potential model, and the bound states are labelled by the principal quantum number $n$, the
orbital angular momentum $L$, the total angular momentum $J$, and the spin multiplicity 
$\sigma$ ($\sigma=1~{\rm or}~ 3$), i.e., $n^{\sigma}L_J$. The typical radius $a=\sqrt{\langle
r^2\rangle}$ of the $c\bar{c}$ and $b\bar{b}$ quarkonia obtained
in this way is of the order of $~10^{-1}$ fm. With such a small
radius, the idea of multipole radiation can be applied to the soft
gluon emissions in hadronic transitions. 
Consider an emitted gluon with a momentum $k$. 
For typical hadronic transition processes, $k\sim{\rm few~hundred~MeV}$, so that $~ak~$ is of the 
order of $10^{-1}$. Thus multipole expansion works for hadronic transition processes. 
Note that the convergence of QCDME does not depend on the value of the QCD coupling constant $g_s$. 
Therefore QCDME is a feasible approach to the soft gluon emissions in hadronic transitions (\ref{HT}).

QCDME has been studied by many authors
\cite{Gottfried,BR,Peskin,VZNS,Yan,KYF}. The gauge invariant
formulation is given in Ref. \cite{Yan}. Let $\psi(x)$ and
$A^a_\mu(x)$ be the quark and gluon fields, respectively.
Following Refs. \cite{Yan}, we introduce 
\begin{eqnarray}                        
&&\Psi(\bm x,t)=U^{-1}(\bm x,t)\psi(x),\nonumber\\
&&\frac{\lambda_a}{2}A^{a\prime}_\mu(\bm x,t)
=U^{-1}(\bm x,t)\frac{\lambda_a}{2}A^a_\mu(x)U(\bm x,t)
-\frac{i}{g_s}U^{-1}(\bm x,t)\partial_\mu U(\bm x,t),
\label{Psi}
\end{eqnarray}
where $U(\bm x,t)$ is defined by \cite{KYF}
\begin{eqnarray}                     
U(\bm x,t)\equiv P\exp\bigg[ig_s\int^{\bm x}_{\bm X}\frac{\lambda_a}{2}{\bm A}^a(\bm x^\prime,t)\cdot
d{\bm x}^\prime\bigg],
\label{U}
\end{eqnarray}
in which $P$ is the path-ordering operation, the line integral is
along the straight-line segment connecting the two ends, $\bm
X\equiv (\bm x_1+\bm x_2)/2$ is the center of mass position of
$Q$ and $\bar{Q}$, and $\bm x$ denotes $\bm{x_1}$ or $\bm{x_2}$. 
With these transformed fields, the part of the
QCD Lagrangian related to the heavy quarks becomes \cite{Yan}
\begin{eqnarray}                     
{\cal L}_Q&=&\int\bar{\Psi}\bigg[\gamma^\mu
\bigg(i\partial_\mu-g_s\frac{\lambda_a}{2}A^{a\prime}_\mu
\bigg)-m\bigg]\Psi d^3x\nonumber\\
&&-\frac{1}{2}\frac{g_s^2}{4\pi}\int\sum^8_{a=0}
\bar{\Psi}(\bm x_1,t)\gamma^0
\frac{\lambda_a}{2}\Psi(\bm x_1,t)
\bigg|\frac{1}{\bm x_1-\bm x_2}\bigg|\bar{\Psi}(\bm x_2,t)
\gamma^0\frac{\lambda_a}{2}\Psi(\bm x_2,t)
d^3x_1d^3x_2,
\label{QCDL}
\end{eqnarray}
where $\lambda_0/2\equiv 1$. Note that the transformed quark field $\Psi(\bm x,t)$ is dressed 
with gluons through $U^{-1}(\bm x,t)$ defined in (\ref{U}). We see from Eq.~(\ref{QCDL}) that the 
dressed quark field $\Psi(\bm x,t)$ serves as the {\it constituent quark} field interacting
via the static Coulomb potential in the potential model. In addition, it is the transformed gluon field
$A^{a\prime}_\mu$ (not the original $A^a_\mu$) that appears in the covariant derivative in 
(\ref{QCDL}). $A^{a\prime}_\mu$ contains non-Abelian contributions through $U(\bm x,t)$. 

Following Ref. \cite{Yan}, we
generalize the Coulomb potential in Eq.~(\ref{QCDL}) to the static
potential including the confining potential in potential models,
and we write down the following effective Lagrangian \cite{Yan}
\begin{eqnarray}                    
{\cal L}^{\rm eff}_Q&=&\int\bar{\Psi}\bigg[\gamma^\mu
\bigg(i\partial_\mu-g_s\frac{\lambda_a}{2}A^{a\prime}_\mu
\bigg)-m\bigg]\Psi d^3x
-\frac{1}{2}\int\sum^8_{a=0}\bar{\Psi}(\bm x_1,t)
\gamma^0\frac{\lambda_a}{2}
\Psi(\bm x_1,t)\bigg[\delta_{a0}V_1(|\bm x_1-\bm x_2|)\nonumber\\
&&
+(1-\delta_{a0})V_2(|\bm x_1-\bm x_2|)\bigg]
\bar{\Psi}(\bm x_2,t)\gamma^0
\frac{\lambda_a}{2}\Psi(\bm x_2,t)d^3x_1d^3x_2,\hspace{2.6cm}
\label{effL}
\end{eqnarray}
where $V_1(|\bm{x_1}-\bm{x_2}|)$ is the static potential (including the
confining potential) between $Q$ and $\bar{Q}$ in the
color-singlet state, and $V_2(|\bm{x_1}-\bm{x_2}|)$ is the static
potential between $Q$ and $\bar{Q}$ in the color-octet state. This
${\cal L}^{\rm eff}_Q$ relates the QCD Lagrangian to the potential
models. 

Now we consider the {\it multipole expansion}. Inside the quarkonium, $|\bm{x-X}|\le a$. So we can 
make an expansion by expanding the gluon field $A^a_\mu(\bm x,t)$ in Taylor series of 
$\bm{x-X}$ at the center of mass position $\bm X$. The Taylor series is an expansion in powers of 
the operators $(\bm{x-X})\bm\cdot\bm\nabla$ and $(\bm{x-X})\bm\times\bm\nabla$ applying to the gluon
field. After operating on the gluon field with the gluon momentum $k$, these operators are of the 
order of $ak$. This is QCDME. It has been shown in Ref. \cite{Yan} that this operation leads to
\begin{eqnarray}                 
&&A^{a\prime}_0(\bm{x},t)=A^{a\prime}_0(\bm X,t)-(\bm x-\bm X)\bm\cdot
{\bm E}^a(\bm{X},t)+\cdots,\label{A0}\label{A0}
\\
&&{\bm A}^{a\prime}(\bm{x},t)=-\frac{1}{2}(\bm x-\bm X)
\bm\times {\bm B}^a(\bm X,t)+\cdots,\hspace{1cm}
\label{Ai}
\end{eqnarray}
where ${\bm E}^a$ and ${\bm B}^a$ are color-electric and
color-magnetic fields, respectively. 

In Ref.~\cite{Yan}, the corresponding Hamiltonian was derived based on the above formulation. This is 
more convenient in using the nonrelativistic perturbation theory. The obtain Hamiltonian is \cite{Yan}
\begin{eqnarray}                    
H^{\rm eff}_{QCD}=H^{(0)}_{QCD}+H^{(1)}_{QCD},
\label{Heff}
\end{eqnarray}
where
\begin{eqnarray}                      
H^{(0)}_{QCD}&=&\int\Psi^\dagger(\bm x_1,t)\Psi(\bm x_1,t)
\hat H\Psi^\dagger(\bm x_2,t)\Psi(\bm x_2,t)d^3x_1d^3x_2
\label{H0}
\end{eqnarray}
with                
\begin{eqnarray}                      
\hat H\equiv -\frac{1}{2m_Q}(\partial_1^2+\partial_2^2+V_1(|\bm x_1-\bm x_2|)+\sum^8_{a=1}
\frac{\lambda_a}{2}\frac{\bar\lambda_a}{2}V_2(|\bm x_1-\bm x_2|)+2m_Q,
\label{Hhat}
\end{eqnarray}
and
\begin{eqnarray}                   
H^{(1)}_{QCD}&=&H_1+H_2,\nonumber\\
&&H_1\equiv Q_a A^a_0(\bm X,t),~~~~~~
H_2\equiv -{\bm d}_a\bm\cdot{\bm E}^a(\bm X,t)-
{\bm m}_a\bm\cdot{\bm B}^a(\bm X,t)+\cdots,\hspace{0.5cm}
\label{H1H2}
\end{eqnarray}
in which
\begin{eqnarray}                
&&Q_a\equiv g_E\int\Psi^\dagger(\bm x,t)\frac{\lambda_a}{2}\Psi(\bm x,t)d^3x,
\label{Q}\\
&&{\bm d}_a\equiv g_E\int(\bm x-\bm X)\Psi^\dagger(\bm x,t)\frac{\lambda_a}{2}\Psi(\bm x,t)
d^3x,
\label{d}\\
&&{\bm m}_a\equiv \frac{g_M}{2}\int(\bm x-\bm X)\bm\times\Psi^\dagger(\bm x,t){\bm \gamma}
\frac{\lambda_a}{2}\Psi(\bm x,t)d^3x
\label{m}
\end{eqnarray}
are the color charge, color-electric dipole moment, and
color-magnetic dipole moment of the $Q\bar{Q}$ system, respectively. Note
that Eq.~(\ref{effL}) is regarded as an effective Lagrangian.
Considering that the heavy quark may have an anomalous magnetic
moment, we have taken in Eqs.~(\ref{Q}), (\ref{d}) and (\ref{m}) the symbols $g_E$ and $g_M$ to denote
the effective coupling constants for the electric and magnetic
multipole gluon emissions, respectively. We shall see later in
Sec. III that taking $\alpha_E$ and $\alpha_M$ as two parameters
is needed phenomenologically.

We are going to take $H^{(0)}_{QCD}$ as the zeroth order
Hamiltonian, and take $H^{(1)}_{QCD}$ as a perturbation. This is
different from the ordinary perturbation theory since
$H^{(0)}_{QCD}$ is not a free field Hamiltonian. $H^{(0)}_{QCD}$ contains
strong interactions in the potentials in $\hat H$, so that the eigenstates of $H^{(0)}_{QCD}$
are bound states rather than free field states. For a given
potential model, the zeroth order solution can be obtained by
solving the Schr\"odinger equation with the given potential.
Moreover, we see from Eqs.~(\ref{Q}), (\ref{d}) and ({\ref{m}) that only $H_2$ in $H^{(1)}_{QCD}$ 
is of $O(ak)$, while $H_1$ is of $O((ak)^0)$. So that we should keep all orders of $H_1$ in 
the perturbation expansion. 

The general formula for the $S$ matrix element between the initial
state $|I\rangle$ and the final state $|F\rangle$ in this
expansion has been given in Ref. \cite{KYF}, which is
\begin{eqnarray}                        
\langle F|S|I\rangle=-i2\pi\delta(E_F+\omega_F-E_I)
\bigg< F~\bigg|H_2\frac{1}{E_I-H^{(0)}_{QCD}
+i\partial_0-H_1}H_2\cdots H_2\frac{1}{E_I-H^{(0)}_{QCD}+i\partial_0-H_1}H_2\bigg|~I\bigg>,
\label{S}
\end{eqnarray}
where $\omega_F$ is the energy of the emitted gluons. This is the basis of the study of hadronic 
transitions in QCDME. Explicit evaluation of the $S$ matrix elements
in various cases will be presented in Sec. III.

\section{Predictions for Hadronic Transitions in the Single-Channel approach}
 
 In this section, we shall show the predictions for hadronic
 transitions rates in the single-channel approach (inclusion of
 coupled-channel contributions will be given in Sec. IV). In this approach, the
 amplitude of hadronic transitions (\ref{HT}) is diagrammatically shown in FIG.~1
 in which there are two complicated vertices: namely, the 
 vertex of {\it multipole
 gluon emissions} ({\bf MGE}) from  the heavy quarks and the vertex of {\it hadronization} ({\bf H})
 describing the conversion of the emitted gluons
 into light hadron(s). The {\bf MGE} vertex is at the scale of the heavy quarkonium, and it depends 
 on the property of the heavy quarkonium. The {\bf H} vertex is at the scale of the light hadron(s),
and is independent of the property of the heavy quarkonium. In the following, we shall treat them 
separately.
\null\vspace{-0.5cm}
\begin{figure}[h]
\includegraphics[width=6.5truecm,clip=true]{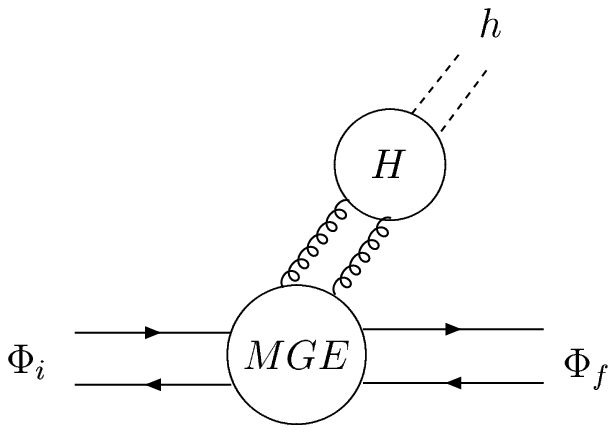}
\null\vspace{-0.4cm}
\caption{\footnotesize Diagram for a typical hadronic transition in the single-channel approach.}
\label{HTfig}
\end{figure}

\subsection{ Hadronic Transitions Between $\bm S$-Wave Quarkonia}

Let us first consider the case of $\pi\pi$ transitions between $S$-wave quarkonia,
$n_I^3S_1\to n_F^3S_1+\pi+\pi$. These processes are dominated by double electric-dipole
transitions (E1E1). The transition amplitude can be obtained from the $S$ matrix element
(\ref{S}). With certain algebra, we obtain \cite{KYF,Yan,KY81}
\begin{eqnarray}                     
{\cal M}_{E1E1}=i\frac{g_E^2}{6}\bigg<\Phi_F h\bigg|{\bm{\bar x}}\cdot{\bm E}
\frac{1}{E_I-H^{(0)}_{QCD}-iD_0}
{\bm{\bar x}}\cdot{\bm E}\bigg|\Phi_I\bigg>,
\label{E1E1}
\end{eqnarray}
where $\bm{\bar x}$ is the separation between $Q$ and $\bar Q$, and 
$(D_0)_{bc}\equiv \delta_{bc}\partial_0-g_s f_{abc}A^a_0$.
Let us insert a complete set of intermediate states with the
principal quantum number $K$ and the orbital angular momentum $L$. 
Then Eq. (\ref{E1E1}) can be written as
\begin{eqnarray}                         
{\cal M}_{E1E1}=i\frac{g_E^2}{6}\sum_{KLK^\prime L^\prime}
\bigg\langle\Phi_F h\bigg|{\bm{\bar x}}\cdot{\bm E}\bigg|K L\bigg
\rangle\bigg\langle K L\bigg|\frac{1}{E_I-H^{(0)}_{QCD}-iD_0}\bigg|K^\prime L^\prime\bigg\rangle
\bigg\langle K^\prime L^\prime\bigg|{\bm{\bar x}}\cdot{\bm E}\bigg|\Phi_I\bigg\rangle,
\label{E1E1'}
\end{eqnarray}
According to the angular momentum selection rule, the intermediate states must have $L=L^\prime=1$.
The intermediate states in the hadronic transition are
the states after the emission of the first gluon and before the
emission of the second gluon shown in FIG.~1, i.e. they are states
with a gluon and a color-octet $Q\bar{Q}$. There is strong interaction between the gluon and the 
color-octet $Q\bar{Q}$ since they all carry colors.
Thus these states are the so-called hybrid states. It is difficult to calculate these hybrid states
from the first principles of QCD. So we shall take a reasonable
model for it. The model should {\it reasonably reflect the main
properties of the hybrid states} and should {\it contain as few
unknown parameters as possible} in order not to affect the
predictive power of the theory. There is a quark confining string
(QCS) model \cite{QCS} satisfying these requirements. The QCS
model is a one dimensional string model in which the strong
confining force between $Q$ and $\bar{Q}$ is described by the
ground-state string, and gluon excitation effects are described by
the vibrations of the string \cite{QCS}. Our intermediate states
$Q\bar{Q}g$ are thus described by the first vibrational mode in
this model. The QCS model is not the only one satisfying the above
requirements. Another possible model satisfying the requirements
is the MIT bag model for the hybrid states. Hadronic transitions
with the MIT bag model as the model for the intermediate states
has been studied in Ref.~\cite{bag}. It is shown that, with the
same input data, the predictions in this model are very close to
those in the QCS model although the absolute intermediate state
energy eigenvalues in the bag model are much higher than those in
the QCS model. Thus the predictions are not sensitive to the 
specific energy spectrum of the intermediate states. 
In the following, we take the calculations with the QCS model as
examples. Explicit calculations of the first vibrational mode as
the intermediate states is given in Ref. \cite{KY81}. With this
model, the transition amplitude (\ref{E1E1'}) becomes \cite{KY81}
\begin{eqnarray}                            
{\cal M}_{E1E1}=i\frac{g_E^2}{6}\sum_{KL}\frac{\langle\Phi_F|\bar x_k|KL\rangle
\langle KL|\bar x_l|\Phi_I\rangle}
{E_I-E_{KL}}\langle \pi\pi|E^a_k E^a_l|0\rangle,
\label{factorizedE1E1}
\end{eqnarray}
where $E_{KL}$ is the energy eigenvalue of the intermediate
vibrational state $|KL\rangle$. We see that, in this approach, the
transition amplitude contains two factors: namely, the heavy quark
{\bf MGE} factor (the summation) and the {\bf H} factor
$\langle\pi\pi|E^a_k E^a_l|0\rangle$. The first factor concerns
the wave functions and energy eigenvalues of the initial and final
state quarkonia and the intermediate states. These can be
calculated for a given potential model. Let us now consider the
treatment of the second factor. The scale of the {\bf H}
factor is the scale of light hadrons which is very low. Therefore
the calculation of this matrix element is highly nonperturbative.
So far, there is no reliable way of calculating this {\bf H} factor 
from the first principles of QCD. Therefore we take
a phenomenological approach based on an analysis of the structure
of this matrix element using PCAC and soft pion technique in Ref.
\cite{BrownCahn}. In the center-of-mass frame, the two pion momenta $q_1$ and $q_2$ are the only 
independent variables describing this matrix element. According Ref. \cite{BrownCahn}, we can write
this matrix element as \cite{KY81}
\begin{eqnarray}                    
\frac{g_E^2}{6}\langle\pi_\alpha(q_1)\pi_\beta(q_2)|E^a_k E^a_l|0\rangle
=\frac{\delta_{\alpha\beta}}{\sqrt{(2\omega_1)(2\omega_2)}}
\bigg[C_1\delta_{kl}q^\mu_1 q_{2\mu}
+C_2\bigg(q_{1k}q_{2l}+q_{1l}q_{2k}-\frac{2}{3}\delta_{kl}{\bm q}_1
\cdot{\bm q}_2\bigg)\bigg],
\label{HofE1E1}
\end{eqnarray}
where $C_1$ and $C_2$ are two unknown constants. In the rest frame of $\Phi_I$ (the center-of-mass 
frame), for a given $\pi\pi$ invariant mass $M_{\pi\pi}$, the $C_1$ term is isotropic
($S$-wave), while the $C_2$ term is angular dependent ($D$-wave).
In the nonrelativistic single-channel approach, the {\bf MGE} factor in (\ref{factorizedE1E1})
is proportional to $\delta_{kl}$ due to orbital angular momentum conservation. So that
only the $C_1$
term contributes to the $S$-state to $S$-state transitions. In
this case, the $n_I^3S_1\to n_F^3S_1+\pi+\pi$ transition rate can
be expressed as \cite{KY81}
\begin{eqnarray}                        
\Gamma(n_I^3S_1\to n_F^3S_1~\pi~\pi)=|C_1|^2G|f^{111}_{n_I0n_F0}|^2,
\label{GammaS-S}
\end{eqnarray}
where the phase-space factor $G$ is \cite{KY81}
\begin{eqnarray}                      
G\equiv\frac{3}{4}\frac{M_{\Phi_F}}{M_{\Phi_I}}\pi^3\int K\sqrt{1-
\frac{4 m_\pi^2}{M_{\pi\pi}^2}}
(M_{\pi\pi}^2-2m_\pi^2)^2~dM_{\pi\pi}^2,
\label{G}
\end{eqnarray}
with
\begin{eqnarray}                      
~~K\equiv \frac{\sqrt{(M_{\Phi_I}
+M_{\Phi_F})^2-M_{\pi\pi}^2}\sqrt{(M_{\Phi_I}-M_{\Phi_F})^2-M_{\pi\pi}^2}}{2M_{\Phi_I}},
\label{K}
\end{eqnarray}
and
\begin{eqnarray}                     
f^{LP_IP_F}_{n_Il_In_Fl_F}\equiv \sum_{K}\frac{\int R_F(r)r^{P_F}R^*_{KL}(r)r^2dr
\int R^*_{KL}(r^\prime)
r^{\prime P_I}R_I(r^\prime)r^{\prime 2}dr^\prime}{M_I-E_{KL}},
\label{f}
\end{eqnarray}
in which $R_I$, $R_F$, and $R_{KL}$ are radial wave functions of
the initial, final, and intermediate vibrational states,
respectively. These radial wave functions are calculated from the 
Schr\"odinger equation with a given potential model.

Now there is only one overall unknown constant $C_1$ left in this
transition amplitude, and it can be determined by taking a
well-measured hadronic transition rate as an input. So far, the
best measured $S$-state to $S$-state $\pi\pi$ transition rate is
$\Gamma(\psi^\prime\to J/\psi~\pi\pi)$. The updated experimental
value is \cite{PDG}
\begin{eqnarray}                       
&&\Gamma_{\rm tot}(\psi^\prime)=281\pm 17~{\rm keV},\nonumber\\
&&B(\psi^\prime\to J/\psi~\pi^+\pi^-)=(31.8\pm 1.1)\%,\nonumber\\
&&B(\psi^\prime\to J/\psi~\pi^0\pi^0)=(18.8\pm 1.2)\%.
\label{input}
\end{eqnarray}
We take this as an input to determine $C_1$. Then we can predict
all the $S$-state to $S$-state $\pi\pi$ transitions rates in the
$\Upsilon$ system. Since the transition rate (\ref{GammaS-S}) depends on
the potential model through the amplitude (\ref{f}), the determined
value of $|C_1|$ is model dependent. In the following, we take the
Cornell Coulomb plus linear potential model \cite{Cornell} and the
Buchm\"uller-Grunberg-Tye (BGT) potential model \cite{BGT} as
examples to show the determined $|C_1|$ and the predicted rates of
$\Upsilon^\prime\to \Upsilon~\pi\pi$, $\Upsilon^{\prime\prime}\to
\Upsilon~\pi\pi$, and
$\Upsilon^{\prime\prime}\to\Upsilon^\prime~\pi\pi$. The results are
listed in TABLE I \footnote{The calculated results are given in
Ref. \cite{KY81}. However, the updated results listed in TABLE I
are larger than those in Ref. \cite{KY81} by approximately a
factor of 1.3 because the updated experimental value of
$\Gamma(\psi^\prime\to J/\psi~\pi\pi)$ is larger than the old
experimental value used in Ref. \cite{KY81} by approximately a
factor of 1.3.}.
\begin{table}[h]
{\small \caption{\footnotesize The determined $|C_1|^2$ and the predicted rates
$\Gamma(\Upsilon^\prime\to \Upsilon~\pi\pi)$,
$\Gamma(\Upsilon^{\prime\prime}\to \Upsilon~\pi\pi)$, and
$\Gamma(\Upsilon^{\prime\prime}\to\Upsilon^\prime~\pi\pi)$ ( in
keV) in the Cornell model and the BGT model. The corresponding
updated experimental values of the transition rates quoted from
Ref. \cite{PDG} are also listed for comparison.} 
\tabcolsep 0.6cm
\vspace{0.4cm}
\begin{tabular}{cccc}
\hline\hline
 &Cornell&BGT&Expt.\\
 \hline
$|C_1|^2\hspace{2.8cm}$&$83.4\times 10^{-6}$&$67.8\times 10^{-6}$&\\
$\Gamma(\Upsilon^\prime\to\Upsilon\pi\pi)$~(keV)~~&8.6~~~&7.8~~~&$~~~12.0~\pm 1.8~~~~~~$\\
$\Gamma(\Upsilon^{\prime\prime}\to\Upsilon\pi\pi)$~(keV)&0.44&1.2~~~&$1.72\pm 0.35$\\
$\Gamma(\Upsilon^{\prime\prime}\to\Upsilon^\prime\pi\pi)$~(keV)&0.78&0.53&$1.26\pm 0.40$\\
\hline\hline
\end{tabular}
}
\end{table}
We see that the predicted ratios
$\Gamma(\Upsilon^{\prime\prime}\to\Upsilon~\pi\pi)/
\Gamma(\Upsilon^\prime\to \Upsilon~\pi\pi)\approx 1.2/7.8=0.15$~
and ~
$\Gamma(\Upsilon^{\prime\prime}\to\Upsilon^\prime~\pi\pi)/\Gamma(\Upsilon^\prime
\to\Upsilon~\pi\pi)\approx 0.53/7.8=0.07$ in the BGT model are
close to the corresponding experimental values
$\Gamma(\Upsilon^{\prime\prime}\to\Upsilon~\pi\pi)/
\Gamma(\Upsilon^\prime\to \Upsilon~\pi\pi)\approx 1.72/12.0=0.14$
and
$\Gamma(\Upsilon^{\prime\prime}\to\Upsilon^\prime~\pi\pi)/\Gamma(\Upsilon^\prime
\to\Upsilon~\pi\pi)\approx 1.26/12.0=0.11$. However, the predicted
absolute partial widths are smaller than the corresponding
experimental values by roughly a factor of $(50\--75)\%$. Moreover, when the
$M_{\pi\pi}$ distributions are considered, the situation will be more complicated.
We shall deal with these issues in Sec. IV.

Note that the phase space factor $G$ in
$\Upsilon^{\prime\prime}\to\Upsilon~\pi\pi$ is much larger than
that in $\Upsilon^\prime\to \Upsilon~\pi\pi$,
$G(\Upsilon^{\prime\prime}\to\Upsilon~\pi\pi)/ G(\Upsilon^\prime\to
\Upsilon~\pi\pi)=33$. So one may naively expect that
$\Gamma(\Upsilon^{\prime\prime}\to\Upsilon~\pi\pi)>\Gamma(\Upsilon^\prime\to
\Upsilon~\pi\pi)$. However, as we see from the experimental values
in TABLE I that $\Gamma(\Upsilon^{\prime\prime}\to\Upsilon~\pi\pi)
/\Gamma(\Upsilon^\prime\to \Upsilon~\pi\pi)\approx 1.72/12.2=0.14$.
The reason why our predictions for this ratio is close to the
experimental value is that the contributions from various
intermediate states to the overlapping integrals in the summation
in $f^{111}_{3010}$ [cf. Eq.~(\ref{f})] {\it drastically cancel}
each other due to the fact that the $\Upsilon^{\prime\prime}$ wave
function contains two nodes. This is a {\it characteristic} of this kind
of intermediate state models (QCS or bag model). To see this, let
us take a simplified model for the intermediate states and look at
its prediction. If we make a simplification assumption that the
variation of the factor $1/(E_I-H^{(0)}_{QCD}-iD_0)$ in Eq.
(\ref{E1E1'}) is sufficiently slow such that it can be approximately
represented by a constant which can be taken out of the summation,
the summations $\displaystyle \sum_{KL}$ and $\displaystyle
\sum_{K^\prime L^\prime}$ in (\ref{E1E1'}) can then be carried out and
the double overlapping integrals in the numerator in (\ref{f})
reduces to a single integration $\int
R^*_{\Upsilon^{\prime\prime}}(r)r^2R_\Upsilon(r)r^2dr$. This
simplified model predicts a rate
$\Gamma(\Upsilon^{\prime\prime}\to\Upsilon~\pi\pi)$ larger than the
experimental value by orders of magnitude \cite{KY81}. Therefore,
taking a reasonable model for the intermediate states is crucial
for obtaining successful predictions in the QCDME approach to the {\bf MGE} factor.

The transitions $~n_I^3S_1\to n_F^3S_1+\eta~$ are contributed
by E1M2 and M1M1 transitions, and is
dominated by the E1M2 transition. The transition amplitude is
\begin{eqnarray}                 
{\cal M}_{E1M2}=-\frac{i}{2m_Q}\frac{g_Eg_M}{6}\sum_{KL}\frac{\langle\Phi_F|\bar x_k|KL\rangle
\langle KL|
S_l\bar x_m|\Phi_I\rangle+\langle\Phi_F|S_l\bar x_m|KL\rangle\langle KL|\bar x_k|\Phi_I\rangle}
{E_I-E_{KL}}
\langle \eta|E^a_k \partial_m B^a_l|0\rangle,
\label{E1M2}
\end{eqnarray}
where $\bm S$ is the total spin of the quarkonium. The {\bf MGE} matrix
element is proportional to $\delta_{km}$. Similar to the idea in
(\ref{HofE1E1}), we can phenomenologically parameterize the hadronization
factor according to its Lorentz structure as \cite{KTY88}
\begin{eqnarray}                    
\frac{g_Eg_M}{6}\langle\eta(q)|E^a_k\partial_k B^a_l|0\rangle=i(2\pi)^{3/2}~C_3~ q_l,
\label{HofE1M2}
\end{eqnarray}
in which the phenomenological constant can be determined by taking the data \cite{PDG}
\begin{eqnarray}                   
\Gamma_{\rm tot}(\psi^\prime)=277\pm22~{\rm keV},
~~~~~~~~B(\psi^\prime\to J/\psi~\eta)=(3.17\pm 0.21)\%
\label{input'}
\end{eqnarray}
as input, and so we can predict the rates for $\Upsilon^\prime\to\Upsilon~\eta$ and
$\Upsilon^{\prime\prime}\to\Upsilon~\eta$. This is equivalent to
\begin{eqnarray}                       
\Gamma(\Upsilon(n_I^3S_1)\to\Upsilon~\eta)=\displaystyle\frac{\bigg|\displaystyle
\frac{f^{111}_{n_I010}(b\bar{b})}{m_b}\bigg|^2}
{\bigg|\displaystyle\frac{f^{111}_{2010}(c\bar{c})}{m_c}\bigg|^2}\frac{|{\bm q}(b\bar{b)}|^3}
{|{\bm q}(c\bar{c})|^3}
\Gamma(\psi^\prime\to J/\psi~\eta).
\label{Upsilon,eta}
\end{eqnarray}
where $\bm q(b\bar{b})$ and $\bm q(c\bar{c})$ are the momenta of $\eta$ in 
$\Upsilon(n_I^3S_1)\to\Upsilon~\eta$ and $\psi^\prime\to J/\psi~\eta$, respectively.
Taking the BGT model as an example to calculate the ratio of transition amplitudes in 
(\ref{Upsilon,eta}), we obtained
\begin{eqnarray}                         
\Gamma(\Upsilon^\prime\to\Upsilon~\eta)=0.022~{\rm keV},~~~~~~
\Gamma(\Upsilon^{\prime\prime}\to\Upsilon~\eta)=0.011~{\rm keV}.
\label{Upsilon'Upsilon",eta}
\end{eqnarray}
These are consistent with the present experimental bounds \cite{PDG}
\begin{eqnarray}                  
\Gamma(\Upsilon^\prime\to\Upsilon~\eta)<0.086~{\rm keV},~~~~~~
\Gamma(\Upsilon^{\prime\prime}\to\Upsilon`~\eta)<0.058~{\rm keV}.
\label{eta-expt}
\end{eqnarray}

We can also compare the ratios 
$R^\prime\equiv \Gamma(\Upsilon^\prime\to\Upsilon~\eta)/\Gamma(\psi^\prime\to J/\psi~\eta)$ and
$R^{\prime\prime}\equiv \Gamma(\Upsilon^{\prime\prime}
\to\Upsilon~\eta)/\Gamma(\psi^\prime\to J/\psi~\eta)$ with the recent experimental measurements.
Recently BES has obtained an accurate measurement of $\Gamma(\psi^\prime\to J/\psi~\eta)$ and 
$\Gamma(\psi^\prime\to J/\psi~\pi^0)$ 
\cite{besggJ}. With the new BES data and the bounds on $\Gamma(\Upsilon^\prime\to\Upsilon~\eta)$ and 
$\Gamma(\Upsilon^{\prime\prime}\to\Upsilon~\eta)$ \cite{PDG}, the experimental bounds on $R^\prime$ and 
$R^{\prime\prime}$ are \cite{besggJ}
\begin{eqnarray}                       
R^\prime|_{expt}<0.0098,~~~~R^{\prime\prime}|_{expt}<0.0065.
\label{R'R"expt}
\end{eqnarray}
Taking the BGT model to calculate the ratios $R^\prime$ and $R^{\prime\prime}$, we obtain
\begin{eqnarray}                    
R^\prime|_{BGT}=
0.0025,~~~~~~~~~~
R^{\prime\prime}|_{BGT}=
0.0013.
\label{R'R"}
\end{eqnarray}
These are consistent with the new experimental bounds (\ref{R'R"expt}).

\subsection{$\bm\pi\bm\pi$ Transitions Between $\bm P$-Wave Quarkonia}

Let us consider the hadronic transitions $2^3 P_{J_I}\to 1^3
P_{J_F}+\pi+\pi$. For simplicity, we use the symbol $\Gamma(J_I\to
J_F)$ to denote $\Gamma(2^3 P_{J_I}\to1^3 P_{J_F}\pi\pi)$. These
are also dominated by E1E1 transitions. The obtained results are
\cite{Yan,KY81}
\begin{eqnarray}                 
&&\Gamma(0\to 0)=\frac{1}{9}|C_1|^2G\bigg|f^{011}_{2111}+2f^{211}_{2111}\bigg|^2,
~~~~\Gamma(0\to 1)=\Gamma(1\to 0)=0,\nonumber\\
&&\Gamma(0\to 2)=5\Gamma(2\to 0)=\frac{10}{27}|C_2|^2H\bigg|f^{011}_{2111}+
\frac{1}{5}f^{211}_{2111}\bigg|^2,
~~~~\Gamma(1\to 1)=\Gamma(0\to 0)+\frac{1}{4}\Gamma(0\to 2),\nonumber\\
&&\Gamma(1\to 2)=\frac{5}{3}\Gamma(2\to 1)=\frac{3}{4}\Gamma(0\to 2),
~~~~\Gamma(2\to 2)=\Gamma(0\to 0)+\frac{7}{20}\Gamma(0\to 2),
\label{P-P}
\end{eqnarray}
where the phase-space factor $H$ is
\begin{eqnarray}                 
H=\frac{1}{20}\frac{M_{\Phi_F}}{M_{\Phi_I}}\pi^3
\int K\sqrt{1-\frac{4m_\pi^2}{M_{\pi\pi}}}\bigg[
(M_{\pi\pi}^2-4m_\pi^2)^2\bigg(1+\frac{2}{3}\frac{K^2}{M_{\pi\pi}^2}\bigg)
+\frac{8K^4}{15M_{\pi\pi}^4}(M_{\pi\pi}^4+2m_\pi^2M_{\pi\pi}^2+6m_\pi^4)\bigg]~dM_{\pi\pi}^2,
\label{H}
\end{eqnarray}
with $K$ defined in Eq. (\ref{K}).

Now the rates in (\ref{P-P}) depend on both $C_1$ and $C_2$. We know that $C_1$ has been determined
by the input (\ref{input}). So far, there is no well
measured hadronic transition rate available for determining the ratio $C_2/C_1$.
At present, to make predictions, we can only take certain
approximation to estimate $C_2/C_1$ theoretically. The approximation
taken in Ref. \cite{KY81} is to assume that the {\bf H} factor 
$\langle\pi\pi|E^a_kE^a_l|0\rangle$ can be approximately expressed as
\begin{eqnarray}                   
\langle\pi\pi|E^a_kE^a_l|0\rangle\propto \langle gg|E^a_kE^a_l|0\rangle,
\label{2gapprox}
\end{eqnarray}
i.e., $\langle\pi\pi|E^a_kE^a_l|0\rangle$
approximately contains a factor $\langle
gg|E^a_kE^a_l|0\rangle$ and another factor describing the
conversion of the two gluons into $\pi\pi$ which is assumed to be
approximately independent of the pion momenta in the hadronic
transitions under consideration. The R.H.S. of Eq.~(\ref{2gapprox}) can be
easily calculated. Comparing the obtained result with the form
(\ref{HofE1E1}), we obtain
\begin{eqnarray}                     
C_2/C_1\approx 3
\label{C2/C1}
\end{eqnarray}
in such an approximation. This is a crude approximation which can only be regarded as 
an order of magnitude estimate. So it is likely that $C_2/C_1\sim O(1)$ rather than
$O(10^{-1})$ or $O(10)$. A reasonable range of $C_2/C_1$ is 
\begin{eqnarray}                   
1\alt C_2/C_1\alt 3.
\label{range-C2/C1}
\end{eqnarray}

With this range of $C_2/C_1$, the obtained transition rates
$\Gamma(J_I\to J_F)$ of $\chi_b(2^3P_{J_I})\to \chi_b(1^3P_{J_F})\pi\pi$ in the
Cornell model \cite{Cornell} and the BGTmodel \cite{BGT}
are listed in TABLE II. 
The relations between different $\Gamma(J_I\to J_F)$
given in Eq.~(\ref{P-P}) reflect the symmetry in the E1E1 multipole
expansion \cite{Yan}, so that experimental tests of these
relations are of special interest. Recently, CLEO reported a preliminary observation of the 
hadronic transitions $\chi_b(2^3P_{J_I})\to \chi_b(1^3P_{J_F})\pi\pi$ for $J_I=J_F=1$ and 2
\cite{Skwarnicki05}. 
In a very recent paper \cite{Cawlfield}, CLEO measured the transition rate,
and the obtained result is 
$\Gamma(\chi_b(2^3P_{J_I})\to \chi_b(1^3P_{J_F})\pi\pi)=(0.83\pm0.22\pm 0.08\pm 0.19)$ keV 
($J_I=J_F=1,~2$) which is 
consistent with the predicted rates $\Gamma(1\to1)$, and $\Gamma(2\to2)$ listed in TABLE~\ref{2P-1P}.
\null\vspace{-0.6cm}
\begin{table}[h]
\caption{Predicted transition rates $\Gamma(J_I\to J_F)$ of 
$\chi_b(2^3P_{J_I})\to \chi_b(1^3P_{J_F})\pi\pi$ with the parameter range (\ref{range-C2/C1})
in the Cornell model \cite{Cornell} and the BGTmodel \cite{BGT}.}
\null\vspace{0.02cm}
\tabcolsep 14pt
\begin{tabular}{cccccc}
\hline\hline
Model&&&$\Gamma(J_I\to J_F)$~(keV)&&\\
 &$\Gamma(0\to 0)$&$\Gamma(0\to 2)$&$\Gamma(1\to 1)$&$\Gamma(1\to 2)$
 &$\Gamma(2\to 2)$\\
 \hline
 Cornell&0.4 &0.004$\--$0.04 &0.4 &0.003$\--$0.03 &0.4\\
 BGT&0.4 &0.002$\--$0.02 &0.4 &0.001$\--$0.01 &0.4\\
 \hline\hline
\end{tabular}
\label{2P-1P}
\end{table}
\subsection{$\bm\pi\bm\pi$ Transitions of $\bm D$-Wave Quarkonia}

$\psi(3770)$ (or $\psi^{\prime\prime}$) is commonly regarded as
essentially the $1D$ state of the charmonium. It lies above the
$D\bar{D}$ threshold, so that it is usually believed that
$\psi(3770)$ mainly decays into the open channel $D\bar{D}$.
Experimental observations show that the directly measured
$\psi(3770)$ production cross section at $e^+e^-$ colliders is
\cite{CrystalBall84,MARKII}
\begin{eqnarray}                    
\sigma(\psi(3770))=7.5\pm0.8~{\rm nb},
\label{sigma(3770)}
\end{eqnarray}
while the $e^+e^-\to\psi(3770)\to D\bar{D}$ cross section is \cite{MARKIII}
\begin{eqnarray}                     
\sigma(\psi(3770)\to D\bar{D})=5.0\pm 0.5~{\rm nb}.
\label{sigma(DDbar)}
\end{eqnarray}
This discrepancy may indicate that there are considerable
non-$D\bar{D}$ decay modes of $\psi(3770)$. One of the possible
non-$D\bar{D}$ decay modes is the hadronic transition
$\psi(3770)\to J/\psi~\pi\pi$. Theoretical studies of hadronic
transitions of the $D$-wave quarkonia have been carried out by
several authors in different approaches leading to quite different
predictions \cite{BLMN,KY81,MoxhayKo, KY90,Kuang02}. In the
following, we briefly review the approach given in Refs.
\cite{KY90,Kuang02}, and compare the predictions with the recent
experimental result and with other approaches.

The measured leptonic width of $\psi(3770)$ is $(0.26\pm 0.04)$
keV \cite{PDG}. If we simply regard $\psi(3770)$ as a pure $1D$
state of charmonium, the predicted leptonic width will be smaller
than the experimental value by an order of magnitude. Therefore
people consider $\psi(3770)$ as a mixture of charmonium states
\cite{KY90,Kuang02,Godfrey}. State mixing is an important
consequence of the coupled-channel theory, especially for states
close to or beyond the open channel threshold. Take a successful
coupled-channel model, the unitary quark model (UQM) \cite{UQM},
as an example. In this model, $\psi(3770)$ is a mixture of many
$S$-wave and $D$-wave states of charmonium, but the main
ingredients are the $\psi(1D)$ and $\psi(2S)$ states. Neglecting
the small ingredients, we can write $\psi^\prime$ and $\psi(3770)$
as
\begin{eqnarray}              
&&\psi^\prime=\psi(2S)\cos\theta+\psi(1D)\sin\theta,\nonumber\\
&&\psi(3770)=-\psi(2S)\sin\theta+\psi(1D)\cos\theta.
\label{mixing}
\end{eqnarray}
The UQM gives $\theta\approx -8^\circ$ \cite{UQM}. Instead of
taking a specific coupled-channel model, we take a
phenomenological approach determining the mixing angle $\theta$ by
fitting the ratio of the leptonic width of $\psi^\prime$ and
$\psi(3770)$. The leptonic widths of $\psi(2S)$ and $\psi(1D)$ are
proportional to the wave function at the origin $\psi_{2S}(0)$ and
the second derivative of the wave function at the origin
$\displaystyle
\frac{5}{\sqrt{2}}\frac{d^2\psi_{1D}(0)/dr^2}{2m_c^2}$,
respectively. Therefore the determination of $\theta$ depends on
the potential model. Here we take two potential models as
illustration: namely, the Cornell potential model \cite{Cornell}
and the improved QCD motivated potential model by Chen and Kuang
(CK) \cite{CK92} which leads to more successful phenomenological
results. The determined values of $\theta$ are
\begin{eqnarray}                    
&&{\rm Cornell}:~~\theta=-10^\circ,\nonumber\\
&&{\rm CK}:~~~~~~~\theta=-12^\circ.
\label{theta}
\end{eqnarray}
These are all consistent with the UQM value. There can also be an
alternative solution with $\theta\sim 30^\circ$, but it is ruled
out by the measured $M_{\pi\pi}$ distribution of $\psi^\prime\to
J/\psi+\pi+\pi$.

This transition is also dominated by E1E1 gluon emission. The
transition rate is \cite{KY90}
\begin{eqnarray}                       
\Gamma\big(\psi(3770)\to J/\psi~\pi\pi\big)=|C_1|^2\bigg[\sin^2\theta ~G(\psi^\prime)
~|f^{111}_{2010}(\psi^\prime)|^2
+\frac{4}{15}\bigg|\frac{C_2}{C_1}\bigg|^2\cos^2\theta ~H(\psi^{\prime\prime})
~|f^{111}_{1210}(\psi^{\prime\prime})|^2\bigg].
\label{3770rate}
\end{eqnarray}

This transition rate depends on the potential model through the amplitudes
$f^{111}_{2010}$, $f^{111}_{1210}$ and the value of $C_2/C_1$. We take the Cornell
model \cite{Cornell} and the CK model \cite{CK92} as examples. Taking the possible 
range for $C_2/C_1$ given in (\ref{range-C2/C1}), we obtain the values of 
$\Gamma\big(\psi(3770)\to J/\psi+\pi^++\pi^-\big)$
listed listed in TABLE III \footnote{The values listed in TABLE III are larger than
those given in Refs.~\cite{KY90,Kuang02} since the updated input
data is larger.}.
\begin{table}[h]
{\small
\caption{\footnotesize The predicted transition rate
$\Gamma(\psi(3770)\to J/\psi+\pi^++\pi^-)$ (in keV) in the Cornell
model and the CK model with the updated input data (\ref{input}).}
\tabcolsep 2.5cm
\vspace{0.4cm}
\begin{tabular}{cc}
\hline\hline
Model&$\Gamma(\psi(3770)\to J/\psi~\pi^+\pi^-)$~(keV)\\
\hline
Cornell& $26\--139$\\
CK& $32\-- 147$\\
\hline\hline
\end{tabular}
}
\end{table}
Note that $S$-$D$ mixing only affects a few percent of the rate, so that the rate is essentially 
$\Gamma(\psi(1D)\to J/\psi~\pi^+\pi^-)$.

Recently, BES has measured the rate $\Gamma(\psi(3770)\to J/\psi+\pi^++\pi^-)$ based on 27.7 pb$^{-1}$
data of $\psi(3770)$. The measured branching ratio is \cite{BES03}
\begin{eqnarray}                       
B\big(\psi(3770)\to J/\psi+\pi^++\pi^-\big)=(0.34\pm 0.14\pm 0.09)\%.
\label{BES3770-BR}
\end{eqnarray}
With the total width \cite{PDG}
\begin{eqnarray}                     
\Gamma_{tot}\big(\psi(3770)\big)=23.6\pm 2.7~{\rm MeV},
\label{3770width}
\end{eqnarray}
the partial width is \cite{BES03}
\begin{eqnarray}                      
\Gamma_{BES}\big(\psi(3770)\to J/\psi+\pi^++\pi^-\big)=80\pm32\pm21~ {\rm keV}.
\label{BES3770-Gamma}
\end{eqnarray}
This is in agreement with the theoretical predictions in TABLE III. 
Taking the BES data
(\ref{BES3770-Gamma}) and Eq.~(\ref{3770rate}) to determine $C_2/C_1$, we obtain
\begin{eqnarray}                      
C_2/C_1=2^{+0.7}_{-1.3}.
\label{BESC2/C1}
\end{eqnarray}
This shows that $C_2/C_1$ is really of $O(1)$.

Very recently, CLEO-c also detected the channel $\psi(3770)\to J/\psi+\pi^++\pi^-$ with higher precision, 
and the measured branching ratio is \cite{CLEO-c05}
\begin{eqnarray}                        
B\big(\psi(3770)\to J/\psi+\pi^++\pi^-\big)=\big(0.214\pm 0.025\pm 0.022\big)\%.
\label{CLEO3770-BR}
\end{eqnarray}
With the $\psi(3770)$ total width (\ref{3770width}), the partial width is
\begin{eqnarray}                       
\Gamma\big(\psi(3770)\to J/\psi+\pi^++\pi^-\big)=50.5\pm 16.9~{\rm keV}.
\label{CLEO3770-Gamma}
\end{eqnarray}
We can also determine $C_2/C_1$ from (\ref{CLEO3770-Gamma}) and (\ref{3770rate}), and the result is
\begin{eqnarray}                        
C_2/C_1=1.52^{+0.35}_{-0.45}.
\label{CLEO-cC2/C1}
\end{eqnarray}
This is consistent with the value (\ref{BESC2/C1}) determined from the BES data, but with 
higher precision.

An alternative way of calculating this kind of transition rate taking the approach to the {\bf H} 
factor proposed by Ref.~\cite{VZNS} was carried out in Ref.~\cite{MoxhayKo}. The so obtained 
transition rate is smaller than the above theoretical prediction by two orders of magnitude. So it
strongly disagrees with (\ref{BES3770-Gamma}) and (\ref{CLEO3770-Gamma}). Therefore
the approach given in Ref.~\cite{VZNS} is ruled out by the BES and CLEO-c experiments.

For the $\Upsilon$ system, state mixings are much smaller
\cite{UQM}. Neglecting state mixings, the $\Upsilon(1D)\to\Upsilon(1S)+\pi^++\pi^-$ transition rate
is proportional to $[C_2/C_1]^2$. Taking the determined values of $C_2/C_1$ in (\ref{BESC2/C1}) and
(\ref{CLEO-cC2/C1}), we obtain the corresponding transition rates: 
$1.3~{\rm keV}\le\Gamma\big(\Upsilon(1D)\to \Upsilon(1S)+\pi^++\pi^-\big)\le 14~{\rm keV}$ 
[from (\ref{BESC2/C1})]
and $2.0~{\rm keV}\le\Gamma(\Upsilon(1D)\to \Upsilon(1S)+\pi^++\pi^-)\le 5.0~{\rm keV}$ 
[from (\ref{CLEO-cC2/C1})], respectively.
The lower values in these ranges are consistent with the CLEO bound \cite{CLEO04}. Improved 
measurement of the $\Upsilon(1D)\to\Upsilon(1S)+\pi^++\pi^-$ rate is desired.

\subsection{Searching for the $\bm h_c$ States}

The spin-singlet $P$-wave states ($1^1P_1$) 
are of special interest since the
difference between the mass of the $1^1P_1$ state and the
center-of-gravity of the $1^3P_J$ states $M_{\rm
c.o.g}=(5M_{1^3P_2}+3M_{1^3P_1}+M_{1^3P_0})/9$ gives useful
information about the spin-dependent interactions between the
heavy quark and antiquark.
There have been various experiments searching for the $h_c$ ($\psi(1^1P_1)$) state. 

In the $\bar{p}p$
collision, $h_c$ can be directly produced. 
In 1992, the E760 Collaboration claimed
seeing a significant enhancement in $\bar{p}p\to J/\psi+\pi^0$ at
$\sqrt{s}=3526.2$ MeV which was supposed to be a candidate of
$h_c$ \cite{E760}. However, such an enhancement has not been
confirmed by the successive E835 experiment from a careful scan in this region with significantly
higher statistics \cite{E835}. Instead, the E835 experiment recently found the $h_c$ state via 
another channel $\bar{p}p\to h_c\to\eta_c\gamma$, and the measured resonance mass is 
$M_{h_c}=3525.8\pm 0.2\pm 0.2$ MeV with a width $\Gamma_{h_c}\alt 1$ MeV \cite{E835}. The measured
production rate is consistent with the theoretical range given in Ref.~\cite{KTY88} 
(see Ref.~\cite{E835}).

At the $e^+e^-$ colliders, the $h_c$ state cannot be produced directly in the $s$-channel due to its
$CP$ quantum number. 
Because of the limited phase space, the best way of 
searching for the $h_c$ state at CLEO-c or BES is through the isospin violating hadronic transition
\cite{CrystalBall83,KTY88,Kuang02}
\begin{eqnarray}                      
\psi^\prime\to h_c+\pi^0.
\label{psi'-h_cpi^0}
\end{eqnarray}
Theoretical calculations of this transition rate considering $S$-$D$ mixing in $\psi'$
and suggestions for tagging the $h_c$ are given in Ref. \cite{Kuang02}. Here we give a brief 
review of it.

The process $\psi^\prime\to h_c+\pi^0$ is dominated by E1M1 transition. The transition amplitude is
\begin{eqnarray}               
{\cal M}_{E1M1}=i\frac{g_Eg_M}{6}\frac{1}{2m_c}\sum_{KL}\frac{\langle h_c|\bar x_k|KL\rangle
\langle KL|(s_c-s_{\bar c})_l|\psi^\prime\rangle+
\langle h_c|(s_c-s_{\bar c})_l|KL\rangle
\langle KL|\bar x_k|\psi^\prime\rangle}{M_{\psi^\prime}-E_{KL}}
\langle \pi^0|E_kB_l|0\rangle,
\label{E1M1amplitude}
\end{eqnarray}
where $\bm s_c$ and $\bm s_{\bar c}$ are spins of $c$ and $\bar{c}$,
respectively. The phenomenological approach to the {\bf H} factor used above does not work in the 
present case since there is no accurate measurement of E1M1 transition rate available as input datum to 
determine the phenomenological parameter so far. Fortunately, evaluation of this special {\bf H} 
factor from QCD turns out to be easy. Since $\pi^0$ is a pseudoscalar, the {\bf H} factor 
$\langle\pi^0|E_kB_l|0\rangle$ is nonvanishing only when $E_kB_l=\delta_{kl}\bm{E}\cdot\bm{B}/3$, and
$\bm{E}\cdot\bm{B}$ is related to the axial-vector anomaly. Therefore
\begin{eqnarray}                 
\langle \pi^0(\eta)|\alpha_s E^a_kB^a_l|0\rangle
=\frac{1}{3}\delta_{kl}\langle\pi^0(\eta)|\alpha_s\bm E^a\cdot\bm B^a|0\rangle
=\frac{1}{12}\delta_{kl}\langle\pi^0(\eta)|\alpha_s F^a_{\mu\nu}\tilde{F}^{a\mu\nu}|0\rangle,
\label{anomaly}
\end{eqnarray}
and the last matrix element can be evaluated by using the Gross-Treiman-Wilczeck
formula \cite{GTW} which leads to
\begin{eqnarray}                     
\langle\pi^0|\alpha_sF^a_{\mu\nu}\tilde{F}^{a\mu\nu}|0\rangle
=\frac{4\pi}{\sqrt{2}}\frac{m_d-m_u}{m_d+m_u}f_\pi m_\pi^2,~~~~~~~
\langle\eta|\alpha_sF^a_{\mu\nu}\tilde{F}^{a\mu\nu}|0\rangle
=\frac{4\pi}{\sqrt{6}}f_\pi m_\eta^2,
\label{GTWresult}
\end{eqnarray}
in which the factor $(m_d-m_u)/(m_d+m_u)$ reflects the violation
of isospin. To predict the transition rate with these expressions, 
we should determine the relations between the effective
coupling constants $\alpha_E=\displaystyle\frac{g^2_E}{4\pi}$,
$\alpha_M=\displaystyle\frac{g^2_M}{4\pi}$ and the coupling constant
$\alpha_s$ appearing in Eqs. (\ref{anomaly}) and (\ref{GTWresult}). With certain
approximations, we can calculate the transition rates
$\Gamma(\psi^\prime\to J/\psi~\pi\pi)$ and $\Gamma(\psi^\prime\to
J/\psi~\eta)$ expressed in terms of $\alpha_E$ and $\alpha_M$
\cite{KY81,KTY88}, so that $\alpha_E$ and $\alpha_M$ can be
determined by taking the input data (\ref{input}) and (\ref{input'})
\footnote{In such an approach, it is not possible to simply take
$\alpha_E=\alpha_M$ to fit the two input data. This is why we take
$\alpha_E$ and $\alpha_M$ as two parameters in our whole approach.
Furthermore, the updated input datum of $\Gamma(\psi^\prime\to
J/\psi~\pi\pi)$ obtained from (\ref{input}) is larger than the old value
used in Ref. \cite{KY81}, so that the determined $\alpha_E$ in Eq.
(\ref{alpha_E}) is larger than the values listed in Ref. \cite{KY81}.}.
The determined $\alpha_E$ is approximately
\begin{eqnarray}                    
\alpha_E\approx 0.6,
\label{alpha_E}
\end{eqnarray}
while the determination of $\alpha_M$ is  quite uncertain because
the approximation used in calculating $\Gamma(\psi^\prime\to
J/\psi~\eta)$ is rather crude \cite{KTY88}. So we take a possible
range \cite{KTY88}
\begin{eqnarray}                     
\alpha_E\le\alpha_M\le 3\alpha_E
\label{alpha_M}
\end{eqnarray}
to estimate the rate. Since the value of $\alpha_E$ in (\ref{alpha_E}) is
just about the commonly estimated value of the strong coupling
constant $\alpha_s$ at the light hadron scale, we simply take
$\alpha_s\approx \alpha_E$. In this spirit, taking account of the $S$-$D$ mixing (\ref{mixing})
in $\psi'$, the transition rate of (\ref{psi'-h_cpi^0}) is 
\begin{eqnarray}                    
\Gamma(\psi^\prime\to h_c\pi^0)&=&\frac{\pi^3}{143m_c^2}\bigg(\frac{\alpha_M}{\alpha_E}\bigg)
\bigg|\cos\theta\bigg(f^{110}_{2011}+f^{001}_{2011}\bigg)-\sqrt{2}\sin\theta
\bigg(f^{110}_{1211}+f^{201}_{1211}\bigg)\bigg|^2\nonumber\\
&&\times\frac{E_{h_c}}{M_{\psi^\prime}}
\bigg[\frac{m_d-m_u}{m_d+m_u}f_\pi m_\pi^2\bigg]^2|\bm q_\pi|.
\label{Gamma(psi'-h_cpi^0)}
\end{eqnarray}
Here we have neglected the state mixing effect in $h_c$ which is
small \cite{UQM} since $h_c$ is not close to the $D\bar{D}$
threshold. Numerical result in the CK potential model is \cite{Kuang02}
\begin{eqnarray}                    
\Gamma(\psi^\prime\to h_c\pi^0)=0.06\bigg(\frac{\alpha_M}{\alpha_E}\bigg)~{\rm keV},~~~~~~
B(\psi^\prime\to h_c\pi^0)=(2.2\pm 0.2)\bigg(\frac{\alpha_M}{\alpha_E}\bigg)\times 10^{-4}.
\label{psi'-h_cpi^0result}
\end{eqnarray}
The calculation shows that the dependence of the transition rate on
the potential model is mild. 

We know that $\pi^0$ decays $99\%$ into two photons. Thus the
signal in (\ref{psi'-h_cpi^0}) is $\psi^\prime\to h_c\gamma\gamma$ with
$M_{\gamma\gamma}=m_{\pi^0}$. If the momenta of the two photons
can be measured with sufficient accuracy, one can look for the
monotonic $M_{\gamma\gamma}$ as the signal. From the branching
ratio in (\ref{psi'-h_cpi^0result}), we see that, taking account of a $10\%$
detection efficiency, hundreds of signal events can be observed
for an accumulation of 10 millions of $\psi^\prime$. The
backgrounds are shown to be either small or can be clearly
excluded \cite{Kuang02}. Once the two photon energies $\omega_1$
and $\omega_2$ are measured, the $h_c$ mass can be extracted from
the relation
$M_{h_c}^2=M_{\psi^\prime}^2+m_{\pi^0}^2-2M_{\psi^\prime}(\omega_1+\omega_2)$.

To have a clearer signal, one can further look at the decay product of $h_c$.
It has been shown that the main decay channel of $h_c$ is $h_c\to\eta_c\gamma$ \cite{Kuang02}.
So that the easiest signal is $\psi'\to h_c\pi^0\to \eta_c\gamma\gamma\gamma$. 
The branching ratio $B(h_c\to\eta_c\gamma)$ depends on the hadronic width of $h_c$. In 
Ref.~\cite{Kuang02},
the hadronic width of $h_c$ was studied both in the conventional perturbative QCD (PQCD) and in
nonrelativistic QCD (NRQCD) approaches. 

We first look at the PQCD result. With the hadronic width 
obtained from PQCD, Ref.~\cite{Kuang02} predicts
\begin{eqnarray}                          
B(h_c\to\eta_c\gamma)=(88\pm 2)\%.
\label{B(h_c-eta_cgamma)}
\end{eqnarray}
Combining (\ref{psi'-h_cpi^0result}) and (\ref{B(h_c-eta_cgamma)}) with the possible range
(\ref{alpha_M}) of the undetermined parameter $\alpha_M/\alpha_E$, we obtain
\begin{eqnarray}                          
{\rm PQCD}:~~~~~~~~~~~~~~~B(\psi'\to h_c\pi^0)\times B(h_c\to\eta_c\gamma)=(1.9\--5.8)\times 10^{-4}.
\label{PQCDBB}
\end{eqnarray}

Signals considering the exclusive hadronic decays modes of $\eta_c$ are also studied in Ref.~\cite{Kuang02}.

Recently, CLEO-c has found the $h_c$ state via the channel 
$\psi'\to h_c\pi^0\to\eta_c\gamma\gamma\gamma$ \cite{Skwarnicki05,CLEOch_c05}. The measured resonance 
mass is $M_{h_c}=3524.4\pm 0.6\pm 0.4$ MeV \cite{Skwarnicki05,CLEOch_c05} which is consistent with 
the E835 result at the $1\sigma$ level. The measured 
$B(\psi'\to h_c\pi^0)\times B(h_c\to\eta_c\gamma)$ is 
\cite{Skwarnicki05,CLEOch_c05}
\begin{eqnarray}                          
{\rm CLEO-c}:~~~~B(\psi'\to h_c\pi^0)\times B(h_c\to\eta_c\gamma)=(3.5\pm 1.0\pm 0.7)\times 10^{-4}
\label{CLEO-cBB}
\end{eqnarray}
which is in good agreement with the above theoretically predicted range (\ref{PQCDBB}). Future improved
measurement with higher precision can serve as an input to determine the unknown parameter 
$\alpha_M/\alpha_E$.

NRQCD predicts a larger hadronic width of $h_c$, so it predicts a smaller branching ratio of 
$h_c\to\eta_c\gamma$, say $B(h_c\to\eta_c\gamma)=(41\pm 3)\%$ \cite{Kuang02} which leads to 
\begin{eqnarray}                     
{\rm NRQCD}:~~~~~~~~~~~~B(\psi'\to h_c\pi^0)\times B(h_c\to\eta_c\gamma)=
(0.9\--2.7)\times 10^{-4}.
\label{NRQCDBB}
\end{eqnarray}
This is also consistent with the CLEO-c result (\ref{CLEO-cBB}) to the present precision. We expect
future CLEO-c experiment with higher precision to test the PQCD and NRQCD approaches. Since the 
{\bf H} factor (\ref{GTWresult}) in this $\psi'\to h_c\pi^0$ process is obtained from the 
Gross-Treiman-Wilczek relation without taking approximations, the agreement between (\ref{PQCDBB}) 
and (\ref{CLEO-cBB}) implies that the above theoretical approach to the {\bf MGE} factor in 
(\ref{E1M1amplitude}) is quite reasonable. 
CLEO-c has also studied some exclusive hadronic channels \cite{Skwarnicki05,CLEOch_c05}. More accurate 
measurement of the branching ratios of these exclusive hadronic channels may also be compared with 
the corresponding predictions in Ref.~\cite{Kuang02} to test PQCD and NRQCD approaches.

\section{Nonrelativistic Coupled-Channel Approach to  Hadronic Transitions}

We know that an excited heavy quarkonium state lying above the open heavy
flavor threshold can decay into a pair of heavy flavor mesons
${\cal D}$ and $\bar{\cal D}$ (${\cal D}$ stands for the $D$ mesons
if the heavy quark is $c$, and stands for the $B$ mesons if the
heavy quark is $b$). This means that there must exist couplings
between $\Phi$, ${\cal D}$ and $\bar{\cal D}$ shown in FIG.~2.
With such couplings taken into account, a complete theory of heavy
quarkonia satisfying the requirement of {\it unitarity} should
include not only the theory describing the discrete states $\Phi$,
but also the theory describing the continuous sector ${\cal D}\bar{\cal D}$ as well.
Such a theory is the so-called {\it coupled-channel theory}.

\begin{center}
\begin{figure}[h]
\includegraphics[width=4truecm,clip=true]{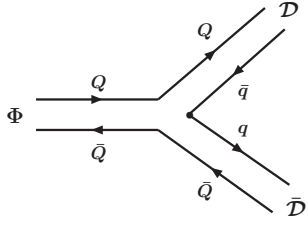}
\caption{\footnotesize Coupling of the heavy quarkonium $\Phi$ to its decay channel 
${\cal D}\bar{\cal D}$.}
\label{Phi-D-D}
\end{figure}
\end{center}

\null\vspace{-1.5cm}

It is hard to study the $\Phi$-${\cal D}$-$\bar{\cal D}$ vertex shown
in FIG.~2 from the first principles of QCD since it is an interaction vertex
between three bound states. There are various models describing
coupled-channel effects, and the two well accepted models are the
Cornell coupled-channel model \cite{Cornell,CCCM} and the UQM \cite{UQM}
mentioned in Sec. III. The $\Phi$-${\cal D}$-$\bar{\cal D}$ vertex
in the UQM is taken to be the $^3P_0$ quark-pair-creation (QPC)
mechanism \cite{QPC}, i.e., the creation of light quark pair $q\bar{q}$ is
supposed to have the vacuum quantum numbers $J^{PC}=0^{++}$ ($^3P_0$),
and the vertex in FIG.~2 is described by the $^3P_0$ sector of the overlapping integral
between the three bound-state wave functions with an almost universal coupling constant
$\gamma_{QPC}\approx 3.03$ \cite{QPC}. The parameters in the UQM are
carefully adjusted so that the model gives good fit to the $c\bar{c}$
and $b\bar{b}$ spectra, leptonic widths, etc. It has been shown that the QPC model
even also gives not bad results for OZI-allowed productions of light
mesons \cite{QPC,QPCappl}, which will be relevant in the calculation of the
hadronic transition amplitudes in FIG.~3(e) and 3(f) in the coupled-channel
theory. So we take the UQM in this section.
\null\vspace{-0.2cm}
\begin{center}
\begin{figure}[h]
\includegraphics[width=10.5truecm,clip=true]{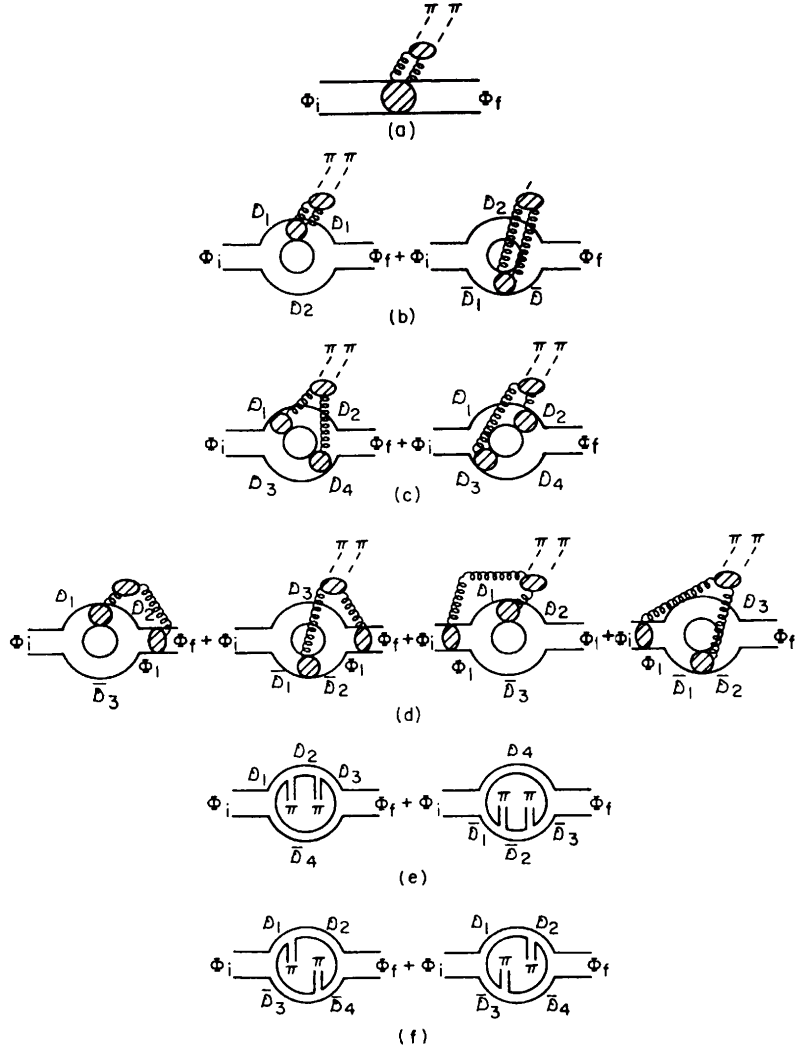}
\vspace{-0.4cm}
\caption{\footnotesize Diagrams for hadronic transitions in the coupled-channel
approach. (e) and (f) are new transitions mechanisms beyond the
QCD multipole expansion. Quoted from Ref.~\cite{ZK91}.}
\label{ZK91FIG1}
\end{figure}
\end{center}

In the UQM, the whole Hilbert space is divided into two sectors:
namely, the confined sector $|\Phi_0;\lambda\rangle$ labelled by
the discrete quantum number $\lambda$ (say $\sigma$, $n$, $L$, $J$) and the continuous sector
$|{\cal D}\bar{\cal D};\nu\rangle$ labelled by the continuous
quantum number $\nu$ (say the momentum). The state
$|\Phi_0;\lambda\rangle$ is just the eigenstate of the Hamiltonian
$H_0$ in the naive single-channel theory with the eigenvalue
$M^0_\lambda$ (the bare mass); i.e.,
\begin{eqnarray}                     
H_0~|\Phi_0;\lambda\rangle=M^0_\lambda~|\Phi_0;\lambda\rangle,
\label{H_0}
\end{eqnarray}
and the state $|{\cal D}\bar{\cal D};\nu\rangle$ is a state with two freely
moving mesons ${\cal D}$ and $\bar{\cal D}$, which is the eigenstate of the
kinetic-energy Hamiltonian $H^c_0$ with the energy eigenvalue $E_\nu$; i.e.,
\begin{eqnarray}                       
H^c_0|{\cal D}\bar{{\cal D}};\nu\rangle=E_\nu|{\cal D}\bar{{\cal D}};\nu\rangle.
\label{H^c_0}
\end{eqnarray}

The total Hamiltonian $H$ of the system contains $H_0$, $H^c_0$ and the
quark-pair-creation Hamiltonian $H_{QPC}$ which determines the OZI-allowed
$\Phi$-${\cal D}$-$\bar{\cal D}$ vertex and mixes the two sectors. $H$ can be
written as
\begin{eqnarray}                      
H=\pmatrix{H_0~~~~ 0\cr 0~~~~ H^c_0}+\pmatrix{~~~0~~~~
H_{QPC}^\dagger\cr H_{QPC}~~~~ 0~~~},
\label{totalH}
\end{eqnarray}
where the first and second rows stand for the confined channel and
the continuous channel, respectively. Note that
\begin{eqnarray}                        
\langle\Phi_0;\lambda|H_{QPC}|\Phi_0;\lambda^\prime\rangle=0,~~~~~~~~
\langle{\cal D}\bar{\cal D};\nu|H_{QPC}|{\cal D}\bar{\cal D};\nu^\prime\rangle=0.
\label{othorgonality}
\end{eqnarray}

\begin{center}
\begin{figure}[h]
\includegraphics[width=5truecm,clip=true]{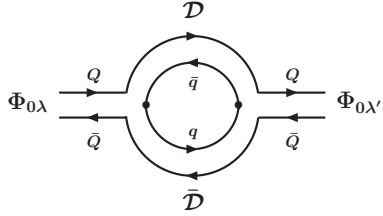}
\vspace{-0.2cm}
\caption{\footnotesize The self-energy $\Pi_{\lambda\lambda^\prime}$ from the ${\cal D}$ meson loop.}
\label{ZK91FIG1}
\end{figure}
\end{center}

\null\vspace{-1cm}
With $H_{QPC}$ introduced, there will be a self-energy $\Pi_{\lambda\lambda^\prime}$ of the 
quarkonium $\Phi_0$ contributed by virtual loops of ${\cal D}$ mesons. This is shown in FIG.~4.
The self-energy $\Pi_{\lambda\lambda^\prime}$ is not necessarily diagonal, i.e., $\lambda$ and
$\lambda^\prime$ may be different. This causes the state mixings. For states below the threshold,
the self-energy $\Pi_{\lambda\lambda^\prime}$ is
\begin{eqnarray}                        
\Pi_{\lambda\lambda^\prime}=-\int
\frac{\langle\Phi_0;\lambda|H_{QPC}|{\cal D}\bar{\cal D};\nu\rangle
\langle{\cal D}\bar{\cal D};\nu|H_{QPC}|\Phi_0;\lambda^\prime\rangle}
{M_\lambda-E_\nu}~d\nu.
\label{Pi}
\end{eqnarray}
Now the total mass matrix of the quarkonium state is
\begin{eqnarray}                        
M_{\lambda\lambda^\prime}=M^0_\lambda\delta_{\lambda\lambda^\prime}+\Pi_{\lambda\lambda^\prime}.
\label{M}
\end{eqnarray}
Let $\alpha_{\lambda\lambda^\prime}$ be the matrix diagonalizing $M_{\lambda\lambda^\prime}$, and
$M_\lambda$ be the diagonal matrix element.
The physical quarkonium state $|\Phi;\lambda\rangle$ is the eigenstate
of $H$ with the energy eigenvalue $M_\lambda$; i.e.,
\begin{eqnarray}                           
H~|\Phi;\lambda\rangle=M_\lambda~|\Phi;\lambda\rangle.
\label{physicalSchrodinger}
\end{eqnarray}
The eigenstate $|\Phi;\lambda\rangle$ can be expressed as a superposition of $|\Phi_0;\lambda\rangle$
and $|{\cal D}\bar{\cal D};\nu\rangle$:
\begin{eqnarray}                            
|\Phi;\lambda\rangle=\sum_\lambda a_{\lambda\lambda^\prime}|\Phi_0;
\lambda^\prime\rangle+\int c_\lambda(\nu)|{\cal D}\bar{\cal D};\nu\rangle~d\nu,
\label{Phi}
\end{eqnarray}
in which all possible open heavy flavor mesons ${\cal D}$
($\bar{\cal D}$) composed of the heavy quark $Q$ and all possible
light quarks $\bar{q}$ ($q$) should be included.

The state mixing coefficient $a_{\lambda\lambda^\prime}$ is related to
$\alpha_{\lambda\lambda^\prime}$ 
by \cite{UQM}
\begin{eqnarray}                       
a_{\lambda\lambda^\prime}=N_\lambda\alpha^T_{\lambda\lambda^\prime},
\label{a-alpha}
\end{eqnarray}
where \cite{UQM}
\begin{eqnarray}                       
\displaystyle
N_\lambda=\left[1+\int\left|\frac{\langle{\cal D}\bar{\cal D};\nu|H_{QPC}|\Phi_0;\lambda\rangle}
{M_\lambda-E_\nu}\right|^2d\nu\right]^{-1/2}
\label{N}
\end{eqnarray}
is a normalization coefficient, and $N_\lambda^2$ determines the
probability of finding the confined sector
$|\Phi_0;\lambda\rangle$ in the physical state
$|\Phi;\lambda\rangle$. The calculation of $\alpha$'s and
$N_\lambda$'s is tedious, and the results for various $c\bar{c}$
and $b\bar{b}$ states are given in Ref. \cite{UQM}.
The mixing coefficient $c_\lambda(\nu)$ is related to $a_{\lambda\lambda^\prime}$ by \cite{UQM}
\begin{eqnarray}                         
c_\lambda(\nu)=\sum_{\lambda^\prime} a_{\lambda\lambda^\prime}\langle{\cal D}\bar{\cal D};\nu|
H_{QPC}|\Phi_0;\lambda^\prime\rangle/(M_\lambda-E_\nu).
\label{c}
\end{eqnarray}

In the single-channel approach, the energy eigenvalues $M^0_\lambda$ of high lying quarkonium 
states predicted by potential models are usually higher than the experimental values. In the 
coupled-channel theory, the self-energy $\Pi_{\lambda\lambda^\prime}$ usually causes 
$M_\lambda<M^0_\lambda$.
Since coupled-channel corrections $M_\lambda-M^0_\lambda$ are unimportant for states lying much lower 
than the ${\cal D}\bar{{\cal D}}$ threshold but are relatively important for states close and above the 
${\cal D}\bar{{\cal D}}$ threshold, coupled-channel theory does improve the prediction for the 
energy spectra.
The UQM coupled-channel theory has been applied to obtain successful
results of heavy quarkonium spectra, leptonic widths, etc., for the
$c\bar{c}$ and $b\bar{b}$ systems \cite{UQM}.

The formulation of the theory of hadronic transitions in the
framework of the UQM was given in Ref. \cite{ZK91}. Let
$H_{pair}\equiv H_{QPC}+H^\dagger_{QPC}$, $\hat{H}_0\equiv H_0$
for the confining sector and $\hat{H}_0\equiv H^c_0$ for the
continuous sector. In the framework of UQM, the $S$ matrix element
(\ref{S}) becomes \cite{ZK91}
\begin{eqnarray}                   
\langle F|S|I\rangle&=&-i2\pi\delta(E_F+\omega_f-E_I)
\bigg< F~\bigg|(H_2+H_{pair})\frac{1}{E_I-\hat{H}_0
+i\partial_0-H_1}(H_2+H_{pair})\cdots\nonumber\\
&&\times
(H_2+H_{pair})\frac{1}{E_I-\hat{H}_0
+i\partial_0-H_1}(H_2+H_{pair})\bigg|~I\bigg>,
\label{CC-S}
\end{eqnarray}

For isospin-conserving $\pi\pi$ transitions (dominated by E1E1
gluon emissions), we take the electric dipole term in $H_2$ [cf.
Eq.~(\ref{H1H2})]. Note that, in Eq. (\ref{CC-S}), the creation of the two
pions can come not only from the conversion of the two emitted
gluons (OZI-forbidden mechanism) via the two $H_2$'s, but also
from the OZI-allowed mechanism $\langle{\cal D}\bar{\cal
D}\pi;\nu|H_{QPC} |{\cal D}\bar{\cal D};\nu\rangle$ directly from
the light quark lines [cf. FIG.~3(e)-3(f)].
Note that the two gluons can only convert into two pions (not one pion) due to
isospin conservation. Thus these two pion creation mechanisms contribute
separately. The $\pi\pi$ transition $S$ matrix element between two
physical quarkonium states $|\Phi;\lambda_I\rangle$ and
$|\Phi;\lambda_F\rangle$ is \cite{ZK91}
\begin{eqnarray}                     
\hspace{-0.8cm}&&\langle\Phi;\lambda_F;\pi({\bm k}_1)\pi({\bm k}_2)|S|\Phi;\lambda_I\rangle=
-i2\pi\delta(M_f+E_{\pi_1}+E_{\pi_2}-M_I)
\sum_{\lambda^\prime_i\lambda^\prime_f}a_{\lambda_I\lambda^\prime_i}
a_{\lambda_F\lambda^\prime_f}
\bigg<\Phi_0;\lambda^\prime_f;\pi({\bm k}_1)\pi({\bm k}_2)\bigg|\nonumber\\
&&\hspace{0.6cm}\times\bigg[H_2\frac{1}{M_I-\hat{H}_0+i\partial_0-H_1}H_2
+H^\dagger_{QPC}\frac{1}{M_I-\hat{H}_0+i\partial_0-H_1}H_2\frac{1}{M_I-\hat{H}_0
+i\partial_0-H_1}H_2\frac{1}{M_I-\hat{H}_0}H_{QPC}\nonumber\\
&&\hspace{0.6cm}+H_2\frac{1}{M_I-\hat{H}_0+i\partial_0-H_1}
H^\dagger_{QPC}\frac{1}{M_I-\hat{H}_0+i\partial_0-H_1}
H_2\frac{1}{M_I-\hat{H}_0}H_{QPC}\nonumber\\
&&\hspace{0.6cm}+H^\dagger_{QPC}\frac{1}{M_I-\hat{H}_0+i\partial_0-H_1}
H_2\frac{1}{M_I-\hat{H}_0+i\partial_0-H_1}
H_{QPC}\frac{1}{M_I-\hat{H}_0+i\partial_0-H_1}H_2\nonumber\\
&&\hspace{0.6cm}+H^\dagger_{QPC}\frac{1}{M_I-\hat{H}_0}H_{QPC}\frac{1}{M_I-\hat{H}_0}H_{QPC}
\frac{1}{M_I-\hat{H}_0}H_{QPC}\bigg]\bigg|\Phi_0;\lambda^\prime_i\bigg>,
\label{CC-S'}
\end{eqnarray}

\null\noindent where $E_{\pi_1}$ and $E_{\pi_2}$ are energies of
the two pions. The Feynman diagrams corresponding to the terms in
(\ref{CC-S'}) are shown in FIG.~3 in whic FIG.~3(a)$\--$Fig.~3(d) are diagrams
corresponding to the first four terms in (\ref{CC-S'}), and FIG.~3(e)$\--$FIG.~3(f) 
are diagrams for the last term in (\ref{CC-S'}). For
convenience, we shall call the first four terms in (\ref{CC-S'}) the
{\bf MGE} part, and call the last term in
(\ref{CC-S'}) the quark-pair-creation ({\bf QPC}) part.

We see that (\ref{CC-S'}) contains much more channels of
$\pi\pi$ transitions than the single-channel theory does. In the
{\bf MGE} part, FIG.~3(a) is similar to FIG.~1 but with state mixings,
so that the single-channel amplitude mentioned in Sec.~IIIA is
only a part of the first term in (\ref{CC-S'}). In the {\bf QPC} part, the
last term in (\ref{CC-S'}) is a {\it new pion creation mechanism through
$H_{QPC}$ irrelevant to {\bf MGE}}. Thus in the
coupled-channel theory, $\pi\pi$ transitions between heavy
quarkonium states are not merely described by QCDME.

Since the state mixings and the QPC vertices depending on the
bound-state wave functions are all different in the $c\bar{c}$ and
the $b\bar{b}$ systems, the predictions for
$\Gamma(\Upsilon^\prime\to \Upsilon~\pi\pi)$,
$\Gamma(\Upsilon^{\prime\prime}\to \Upsilon~\pi\pi)$ and
$\Gamma(\Upsilon^{\prime\prime}\to \Upsilon^\prime~\pi\pi)$ by
taking $\Gamma_{expt}(\psi^\prime\to J/\psi~\pi\pi)$ as input will
be different from those in the single-channel theory. Such
predictions were studied in Ref. \cite{ZK91} in which the same
potential model as in Ref. \cite{UQM} is taken for avoiding the
tedious calculation of $\alpha$'s and $N_\lambda$'s. Note that for
a given QPC model, the {\bf QPC} part in (\ref{CC-S'}) is fixed, while the {\bf MGE}
part still contains an unknown parameter $C_1$ in its
hadronization factor after taking the approximation (\ref{C2/C1}). Since there is 
interference between the {\bf MGE}
part and the {\bf QPC} part, the phase of $C_1$ will affect the result.
Let
\begin{eqnarray}                      
C_1=|C_1|~\displaystyle e^{i\vartheta}.
\label{C_1phase}
\end{eqnarray}
Two input data are thus needed to determine $|C_1|$ and $\vartheta$.
In Ref. \cite{ZK91}, the data of the transition rate and $M_{\pi\pi}$
distribution in $\psi^\prime\to J/\psi~\pi\pi$ are taken
as the inputs. Considering the experimental errors in the $M_{\pi\pi}$
distribution, $\vartheta$ is restricted in the range
~$-1\le \cos\vartheta\le-0.676$. The details of the calculation are
given in Ref. \cite{ZK91} in which the ${\cal D}$ meson states $D~(B)$,
$D^*~(B^*)$, and $D^{**}~(B^{**})$ are taken into account. The so predicted
transition rates
$\Gamma(\Upsilon^\prime\to \Upsilon~\pi\pi)$,
$\Gamma(\Upsilon^{\prime\prime}\to \Upsilon~\pi\pi)$, and
$\Gamma(\Upsilon^{\prime\prime}\to\Upsilon^\prime~\pi\pi)$ for
$\cos\vartheta=-1$ and $\cos\vartheta=-0.676$ are listed in TABLE IV
together with the updated experimental results for comparison.
\begin{table}[h]
{\small \caption{\footnotesize The predicted rates
$\Gamma(\Upsilon^\prime\to \Upsilon~\pi\pi)$,
$\Gamma(\Upsilon^{\prime\prime}\to \Upsilon~\pi\pi)$, and
$\Gamma(\Upsilon^{\prime\prime}\to\Upsilon^\prime~\pi\pi)$ ( in
keV) in the coupled-channel theory with $\cos\vartheta=-1$ and
$\cos\vartheta=-0.676$. The corresponding
updated experimental values of the transition rates quoted from
Ref. \cite{PDG} are also listed for comparison.}
\tabcolsep 0.4cm
\vspace{0.4cm}
\begin{tabular}{ccccc}
\hline\hline
 &&Theory&&Expt.\\
 &$\cos\vartheta=-1$&&$\cos\vartheta=-0.676$&\\
 \hline
$\Gamma(\Upsilon^\prime\to\Upsilon~\pi\pi)$~(keV)~~&14~~~&&13~~~&$~~~12.0~\pm 1.8~~~~~~$\\
$\Gamma(\Upsilon^{\prime\prime}\to\Upsilon~\pi\pi)$~(keV)&~1.1&&~1.0&$1.72\pm 0.35$\\
$\Gamma(\Upsilon^{\prime\prime}\to\Upsilon^\prime~\pi\pi)$~(keV)&~0.1&&~~0.3&$1.26\pm 0.40$\\
\hline\hline
\end{tabular}
}
\end{table}

\null\vspace{-0.5cm}\noindent
We see that the obtained $\Gamma(\Upsilon^\prime\to \Upsilon~\pi\pi)$ is in 
good agreement with the experiment, and the results of
$\Gamma(\Upsilon^{\prime\prime}\to \Upsilon~\pi\pi)$ and 
$\Gamma(\Upsilon^{\prime\prime}\to \Upsilon^\prime~\pi\pi)$ are in 
agreement with the  experiments at the level of $2\sigma$ and 
$2.4\sigma$, respectively.
\begin{center}
\begin{figure}[h]
\includegraphics[width=6.5truecm,clip=true]{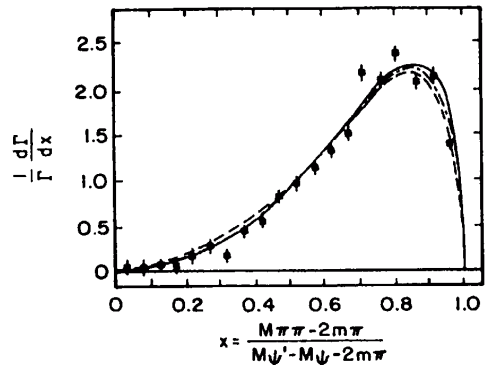}
\vspace{-0.5cm}
\caption{\footnotesize Comparison of the coupled-channel theory predicted
curve of  $d\Gamma(\Upsilon^{\prime}
\to\Upsilon~\pi\pi)/dM_{\pi\pi}$ with the ARGUS data \cite{ARGUS}.
The solid and dashed-dotted lines stand for $\cos\theta=-1$ and
$\cos\theta=-0.676$, respectively. The dashed line is the naive
single-channel result for comparison. Quoted from
Ref.~\cite{ZK91}.} \label{ZK91FIG1}
\end{figure}
\end{center}

\null\vspace{-1.5cm}
Next we look at the predicted $M_{\pi\pi}$ distributions. It is
pointed out in Ref. \cite{ARGUS} that there is a tiny difference
between the the measured $M_{\pi\pi}$ distributions in
$\psi^\prime\to J/\psi~\pi\pi$ and $\Upsilon^\prime\to
\Upsilon~\pi\pi$. In the single-channel theory, the formulas for
these $M_{\pi\pi}$ distributions are the same with the same value
of $|C_1|$. Ref. \cite{ARGUS} tried to explain the tiny difference
by taking the approach to the {\bf H} factor given in Ref. \cite{VZNS}
in which there is a parameter $\kappa$ which is supposed to run. However, the running of
$\kappa$ is not known theoretically, so that it is not clear
whether the running of $\kappa$ from the scale
$M_{\psi^\prime}-M_{J/\psi}=590$ MeV to the scale
$M_{\Upsilon^\prime}-M_\Upsilon=560$ MeV can really explain the
tiny difference or not. Furthermore, as we have seen in Sec.~IIIC that the approach given in
Ref. \cite{VZNS} is ruled out by the recent BES and CLEO-c 
experiments. In the present coupled-channel theory,
once the values of $|C_1|$ and $\vartheta$ are determined by the
input data of $\psi^\prime\to J/\psi~\pi\pi$, the $M_{\pi\pi}$
distribution of $\Upsilon^\prime\to\Upsilon~\pi\pi$ is definitely
predicted. The comparison of the predicted distribution with the
experimental data given in Ref. \cite{ARGUS} is shown in FIG.~5.
We see that the agreement is good, so that the coupled-channel
theory successfully predicts the tiny difference.

However, the situation of the $M_{\pi\pi}$ distributions of
$\Upsilon^{\prime\prime} \to\Upsilon~\pi^+\pi^-$ and
$\Upsilon^{\prime\prime}\to\Upsilon^\prime~\pi^+\pi^-$ are more
complicated. The single-channel theory predicts $M_{\pi\pi}$
distributions similar to FIG.~5 for these two process, i.e., the
distributions are peaked at the large $M_{\pi\pi}$ region. The
CLEO data shows a clear double-peaked shape for the $M_{\pi\pi}$
distribution of $\Upsilon^{\prime\prime}\to\Upsilon~\pi^+\pi^-$
[cf. Fig.~6(a)] \cite{CLEO87,CLEO9491}. The coupled-channel theory
does enhance the low-$M_{\pi\pi}$ region a little, but is far from
giving a double-peaked shape as is shown by the solid and dashed-dottd curves 
in FIG.~6(a).  Actually,
this situation is not only for the coupled-channel theory based on
the UQM. The Cornell coupled-channel model is not substantially
different from the UQM \cite{ZK91}. Compared with the UQM, the
Cornell coupled-channel model leads to relatively larger $S-S$
mixings but smaller $S-D$ mixings after taking the same experimental inputs.
So that the Cornell coupled-channel
model gives even smaller enhancement in the low-$M_{\pi\pi}$ region.
Thus the transition $\Upsilon^{\prime\prime}\to\Upsilon~\pi^+\pi^-$ needs
further investigations with new ideas although the predicted transition rate
$\Gamma(\Upsilon^{\prime\prime}\to\Upsilon~\pi^+\pi^-)$ is consistent
with the CLEO data at the $2\sigma$ level.

\begin{center}
\begin{figure}[h]
\includegraphics[width=11truecm,clip=true]{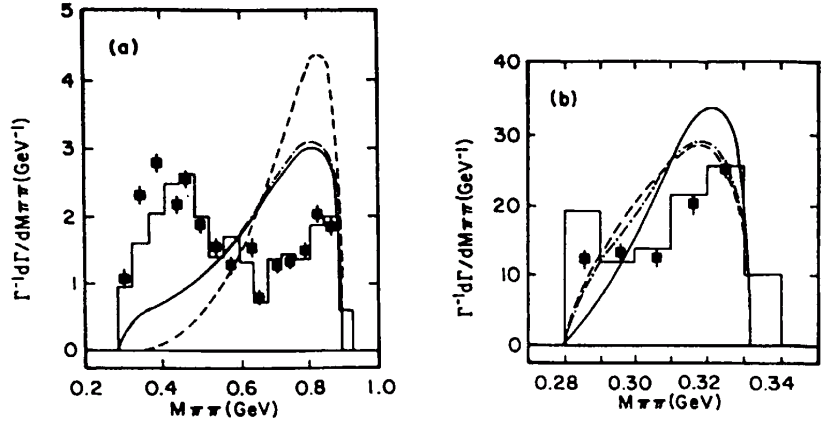}
\caption{\footnotesize Comparison of the coupled-channel theory predicted
curves of $d\Gamma(\Upsilon^{\prime\prime}
\to\Upsilon~\pi^+\pi^-)/dM_{\pi\pi}$ [FIG.~6(a)] and
$d\Gamma(\Upsilon^{\prime\prime}
\to\Upsilon^\prime~\pi^+\pi^-)/dM_{\pi\pi}$ [FIG.~6(b)] with the
CLEO data \cite{CLEO87}. The solid and dashed-dotted lines stand
for $\cos\theta=-1$ and $\cos\theta=-0.676$, respectively. The
dashed line is the naive single-channel result for comparison.
Quoted from Ref.~\cite{ZK91}.} \label{ZK91FIG1}
\end{figure}
\end{center}

\vspace{-1cm} 
There have been various attempts to explain the
double-peaked shape. Ref. \cite{Voloshin-Truong} assumed the
existence of a four-quark state $\Upsilon_1$ having nearly the
same mass as $\Upsilon^{\prime\prime}$ and coupling strongly to
$\Upsilon^{\prime\prime}\pi$ and $\Upsilon\pi$, and the dominant
transition mechanism is suggested to be
$\Upsilon^{\prime\prime}\to \Upsilon_1+\pi\to\Upsilon+\pi+\pi$
which enhances the low-$M_{\pi\pi}$ distribution. The branching
ratio of $\Upsilon(4S)\to \Upsilon_1+\pi$ is estimated to be
roughly $1\%$, so that the assumption can be experimentally tested
by searching for the $\Upsilon_1$ state in $\Upsilon(4S)$ decays.
This idea was carefully studied in Ref. \cite{BDM} taking account
of the final-state $\pi\pi$ interactions and got a double-peaked
shape but the low-$M_{\pi\pi}$ peak is not at the desired
position. A slightly modified model of this kind was proposed in
Ref. \cite{ABSZ}. So far the assumed four-quark state is not
found experimentally. Another attempt was made in Ref.
\cite{Lipkin-Tuan-Moxhay} assuming that the coupled-channel
contributions are strong enough in $\Upsilon^{\prime\prime}\to
\Upsilon~\pi\pi$ that there is a considerably large {\bf QPC} part in the
transition amplitude, and the interference between it and the {\bf MGE}
part may form a double-peaked shape by adjusting the strength of
the {\bf QPC} part. However, as we mentioned above that the strength of
the {\bf QPC} part is fixed once a QPC model is given, and the systematic
calculation in Ref.~\cite{ZK91} shows that the {\bf QPC} part is actually much
smaller than what was expected in Ref. \cite{Lipkin-Tuan-Moxhay}.
Recently, attempts to explain the double-peaked shape by certain
models for a light $\sigma$ meson resonance at around 500 MeV in
the final state $\pi\pi$ interactions with \cite{Ishida01} and
without \cite{Uehara} using the Breit-Wigner formula have been
proposed. By adjusting the free parameters in the
models, the CLEO data on the $M_{\pi\pi}$ distributions in
$\Upsilon^{\prime\prime}\to \Upsilon~\pi\pi$ and
$\Upsilon^{\prime\prime}\to \Upsilon^\prime~\pi\pi$ can be fitted.
However, the models need to be tested in other processes.
Therefore, the $\Upsilon^{\prime\prime}\to \Upsilon~\pi\pi$
transition is still an interesting process needing further
investigations.

We would like to mention that the calculations mentioned above concerns
the wave functions of some excited states of heavy quarkonia, the
heavy flavored mesons ${\cal D}$, and the pions. Nonrelativistic
potential model calculations of these wave functions may not be so
good. Therefore the nonrelativistic coupled-channel
theory of hadronic transitions in Ref. \cite{ZK91} still needs
further improvements. 

\null\noindent
\section{Application of QCD Multipole Expansion to Radiative
Decays of $\bm {J/\psi}$}

In the preceding sections, QCD multipole expansion is applied to
various hadronic transition processes in which the initial- and
final-state quarkonia $\Phi_I$ and $\Phi_F$ are composed of the
same heavy quarks. In this case, the dressed (constituent) quark
field $\Psi(\bm x,t)$ does not actually need to be quantized. Now
we generalize the QCDME theory to processes
including heavy quark flavor changing and heavy quark pair
annihilation or creation. Then the quantization of the $\Psi(\bm
x,t)$ is needed. This has been studied in Ref. \cite{KYF}, and the
obtained canonical commutation relation is \cite{KYF}
\begin{eqnarray}                        
[\Psi(\bm x,t),\Psi^\dagger(\bm x^\prime,t)]=\delta^3(\bm x-\bm x^\prime).
\label{commutator}
\end{eqnarray}
To include the electromagnetic and weak interactions, we generalize the Hamiltonian as
\begin{eqnarray}                          
&&H=H^{(0)}_{QCD}+H_{int},\\
&&H_{int}=H^{(1)}_{QCD}+H_{em}+H_W,
\label{generalH}
\end{eqnarray}
in which $H^{(0)}_{QCD}$ and $H^{(1)}_{QCD}$ are defined in Eqs.
(\ref{H0}) and (\ref{H1H2}), and
\begin{eqnarray}                     
H_{em}&=&e\int d^3x~\bar\Psi(\bm x,t)\gamma^\mu{\cal Q}{\cal A}_\mu(\bm x,t)
\Psi(\bm x,t),\nonumber\\
H_W&=&\int d^3x\bigg[\frac{g}{\sqrt 2}\bar\Psi(\bm x,t)
\gamma^\mu\frac{1-\gamma_5}{2}[t_+W^+_\mu(x)+t_-W^-_\mu(x)]\Psi(\bm x,t)\nonumber\\
&&+\frac{g}{\cos\theta_W}\bar\Psi(\bm x,t)\gamma^\mu
\bigg(\frac{1-\gamma_5}{2}t_3-\sin^2\theta_W{\cal Q}\bigg)Z_\mu(x)\Psi(\bm x,t)\bigg],
\label{Hem-HW}
\end{eqnarray}
where ${\cal Q}$ is the electric charge operator of the heavy
quark, ${\cal A}_\mu$ is the photon field, $g$ and $t_i$ are,
respectively, the weak $SU(2)$ coupling constant and generator,
$\theta_W$ is the Weinberg angle, and $e=g\sin\theta_W$ is the
electromagnetic coupling constant.
\null\vspace{-0.5cm}
\begin{center}
\begin{figure}[h]
\includegraphics[width=11truecm,clip=true]{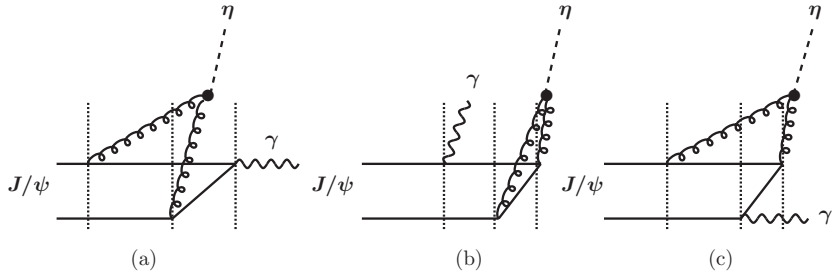}
\caption{\footnotesize Feynman diagrams for the radiative decay process $J/\psi\to\gamma+\eta$. The
intermediate states between two vertical dotted lines are all bound states.}
\end{figure}
\end{center}

\null\vspace{-1.5cm}
Let us take the application of the generalized theory to the radiative decay process 
$J/\psi\to\gamma+\eta$ as an example. 
This process has been studied in the
framework of perturbative QCD and nonrelativistic quark model in
Ref. \cite{KKKS}, but the predicted rate is significantly smaller
than the experimental value. We know that the momentum of the $\eta$ meson in
this process is $q_\eta=(M^2_{L/\psi}-m^2_\eta)/(2M_{J/\psi})=1.5$
GeV. Suppose the $\eta$ meson is converted from two emitted gluons
from the heavy quark. The typical momentum of a gluon is then
$k\sim q_\eta/2\sim 750$ MeV. This is the momentum scale that
perturbative QCD does not work well but QCDME
works \cite{KYF}. So we can calculated the rate of this decay
process using QCDME. The Feynman diagrams for
this process are shown in FIG.~7, in which the intermediate states
marked between two vertical dotted lines are all treated as {\it bound
states} in this approach. In this sense this approach is
nonperturbative, and also for this reason the contributions of the
three diagrams in FIG.~7 are different.

In QCDME, this process is dominated by the E1M2
gluon emissions. So the {\bf H} factor (conversion of the
two gluons into $\eta$) is
\begin{eqnarray}                     
g_Eg_M\langle\eta|E^a_jD_jB^a_i|0\rangle,
\label{E1M2}
\end{eqnarray}
where $D_j\equiv \partial_j-g_s(\lambda_a/2)A^a_j$ is the covariant derivative.
The operator in (\ref{E1M2}) can be written as
\begin{eqnarray}
E^a_jD_jB^a_i=\partial_j(E^a_jB^a_i)-(D_jE^a_j)B^a_i.\nonumber
\end{eqnarray}
It is argued in Ref. \cite{VZNS} that the second term is smaller
than the first term, so they suggested the approximation
\begin{eqnarray}                   
g_Eg_M\langle\eta|E^a_jD_jB^a_i|0\rangle\simeq iq_{\eta j}\langle\eta|E^a_jB^a_i|0\rangle.
\label{VZapprox}
\end{eqnarray}
This matrix element can then be evaluated by using the
Gross-Treiman-Wilczeck formula \cite{GTW}
and we obtain \cite{KYF}
\begin{eqnarray}               
g_Eg_M\langle\eta|E^a_jD_jB^a_i|0\rangle\simeq i\frac{g_Eg_M}{g_s^2}\frac{4\pi^2}{3\sqrt{6}}
q_{\eta i}f_\pi m_\eta^2(\cos\theta_P-\sqrt{2}\sin\theta_P)\delta_{ij},
\label{approxGTW}
\end{eqnarray}
where $\theta_P$ is the mixing angle in the pseudoscalar nonet, i.e.,
\begin{eqnarray}                      
\eta=\eta_8\cos\theta_P-\eta_1\sin\theta_P,~~~~~~\eta^\prime=\eta_8\sin\theta_P+\eta_1\cos\theta_P,
\label{theta_P}
\end{eqnarray}
with
\begin{eqnarray}                      
\eta_8=\frac{1}{\sqrt{6}}(\bar{u}\gamma_5 u+\bar{d}\gamma_5 d-2\bar{s}\gamma_5 s),~~~~
\eta_1=\frac{1}{\sqrt{3}}(\bar{u}\gamma_5 u+\bar{d}\gamma_5 d+\bar{s}\gamma_5 s).
\label{eta8-eta1}
\end{eqnarray}

As in Sec.~IIID, we take $\alpha_s\simeq\alpha_E=0.6$. It is shown in Ref. \cite{KYF} that the
contribution of FIG.~7(a) is larger than those of FIG.~7(b) and FIG.~7(c). Thus as an
approximation, we only consider the main contribution of FIG.~7(a).

The calculated $~J/\psi\to\gamma\eta~$ rate is \cite{KYF}
\begin{eqnarray}                   
\Gamma(J/\psi\to\gamma\eta)=\frac{1}{6\pi}\bigg(\frac{\alpha_M}{\alpha_E}\bigg)
\frac{|\bm q_\eta|^3}{M_{J/\psi}}
\bigg(\frac{2e{\cal Q}}{3\sqrt{6}m_c}\bigg)^2
\bigg[\frac{4\pi^2}{3\sqrt 6}f_\pi m_\eta^2
(\cos\theta_P-\sqrt{2}\sin\theta_P)
\sum_n h^{111}_{10n0}\bigg]^2,
\label{psi-gammaeta}
\end{eqnarray}
where
\begin{eqnarray}                     
h^{LP_iP}_{n_Il_i nl}\equiv \sum_K\frac{\langle R_{n_F}|r^P|R^\prime_{KL}\rangle
\langle R^\prime_{KL}|r^{P_i}|R_{n_Il_i}\rangle}{(M_{J/\psi}-E_{nl}-\omega_\eta)
(M_{J/\psi}-E^\prime_{KL})}f_{nl}(0),
\label{h}
\end{eqnarray}
in which $R_{n_Il_i}$, $R_{nl}$ and $R^\prime_{KL}$ are radial
wave functions of the initial-, final-, and
intermediate-quarkonium states in FIG.~7(a), respectively.
$f_{nl}(0)$ is the wave function at the origin of the
final-quarkonium state. We shall take into account the first five
terms in the summation $\displaystyle\sum_n h^{111}_{10n0}$. As is
well-known that $f_{n0}(0)$ can be determined by the datum of the
related leptonic width $\Gamma(\psi(n^3S_1)\to e^+e^-)$. For
$n=1,2$, the so determined $f_{10}(0)$ and $f_{20}(0)$ are smaller
than the ones predicted by the Cornell potential model by almost
the same factor 0.57 \cite{Cornell}. It is expected that this discrepancy may
be explained by QCD corrections. For $n\ge 3$, the states are
above the threshold and state mixings will be significant, so that
the data of $\Gamma(\psi(n^3S_1)\to e^+e^-)$ are not useful. We
expect that QCD corrections will not vary seriously with $n$ as is
inspired by the cases of $n=1,2$. Then we can calculate
$f_{10}(0),\cdots f_{50}(0)$ using the Cornell potential model,
and then multiply the obtained results by the same factor 0.57 to
obtain the correct values of them. 

The factor $(\cos\theta_P-\sqrt{2}\sin\theta_P)$ in (\ref{approxGTW}) concerns the effective 
$\eta$-$g$-$g$ vertex in the hadronization $gg\to\eta$. This is somewhat similar to the effective 
$\eta$-$\gamma$-$\gamma$ vertex in $\eta\to \gamma\gamma$. We may take the determined value 
of $\theta_P$ from the $\eta\to\gamma\gamma$ and $\eta^\prime\to\gamma\gamma$ data, which is 
$\theta_P\approx -20^\circ$ \cite{PDG}. With this value of $\theta_P$, we get
\begin{eqnarray}                    
\Gamma(J/\psi\to\gamma\eta)=0.041\bigg(\frac{\alpha_M}{\alpha_E}\bigg)~{\rm keV}.
\label{resultpsi-gammaeta}
\end{eqnarray}
With the experimental datum $\Gamma_{tot}(J/\psi)=91.0\pm3.2$ keV \cite{PDG}, we obtain
\begin{eqnarray}                   
B(J/\psi\to\gamma\eta)=(4.5\pm 0.2)\times 10^{-4}\bigg(\frac{\alpha_M}{\alpha_E}\bigg).
\label{B(psi-gammaeta)}
\end{eqnarray}
The experimental value of this branching ratio is \cite{PDG}
\begin{eqnarray}                    
B(J/\psi\to\gamma\eta)\bigg|_{expt}=(8.6\pm 0.8)\times 10^{-4}.\nonumber
\end{eqnarray}
We see that for $\alpha_M/\alpha_E\approx 1.9$, the predicted branching ratio 
agrees with the experimental value.

Note that the value of $\alpha_M/\alpha_E$ and $\theta_P$ are not so certain, and we do not
know how good the approximation (\ref{VZapprox}) really is.
To avoid these uncertainties, we can take the
ratio of $\Gamma(J/\psi\to\gamma\eta)$ to another E1M2 transition
rate $\Gamma(\psi^\prime\to J/\psi\eta)$. The theoretical
prediction is \cite{KYF}
\begin{eqnarray}                       
R_\eta\equiv \frac{\Gamma(J/\psi\to\gamma\eta)}{\Gamma(\psi^\prime\to J/\psi\eta)}
=\frac{\displaystyle\frac{8}{81}(e{\cal Q})^2
|\bm q_\eta(J/\psi\to\gamma\eta)|^3/M_{J/\psi}
\bigg|\displaystyle\sum_n h^{111}_{10n0}\bigg|^2}{\frac{2}{243}
|\bm q_\eta(\psi^\prime\to J/\psi\eta)|^3|f^{111}_{2010}|^2}
=0.012.
\label{R_eta}
\end{eqnarray}
In this ratio, the uncertainties mentioned above are all cancelled, so that $R_\eta$ just tests
the {\bf MGE} mechanism in this approach. The corresponding experimental value is \cite{PDG}
\begin{eqnarray}                        
R_\eta\bigg|_{expt}=0.009\pm 0.003.
\label{R_eta(expt)}
\end{eqnarray}
We see that the agreement is at the $1\sigma$ level. Since we have seen in Sec. IIID that the 
calculation of the {\bf MGE} factor mentioned in Sec.~III is quite reasonable, the agreement of
(\ref{R_eta}) with (\ref{R_eta(expt)}) implies that {\bf MGE} mechanism for this radiative decay 
process is also reasonable.

The above approach can also be applied to the radiative decay process $J/\psi\to\gamma\eta^\prime$.
From (\ref{theta_P}) we see that the $J/\psi\to\gamma\eta^\prime$ decay rate is
\begin{eqnarray}                        
\Gamma(J/\psi\to\gamma\eta^\prime)=\frac{1}{6\pi}\bigg(\frac{\alpha_M}{\alpha_E}\bigg)
\frac{|\bm q_{\eta^\prime}|^3}{M_{J/\psi}}
\bigg(\frac{2e{\cal Q}}{3\sqrt{6}m_c}\bigg)^2\bigg[\frac{4\pi^2}{3\sqrt 6}f_\pi m_{\eta^\prime}^2
(\sqrt{2}\cos\theta_P+\sin\theta_P)
\sum_n h^{111}_{10n0}\bigg]^2,
\label{psi-gammaeta'}
\end{eqnarray}
Since there is no $\psi^\prime\to J/\psi\eta^\prime$ available (not enough phase space), we cannot
have a ratio similar to $R_\eta$ which exactly tests the {\bf MGE} mechanism. We can define the ratio
\begin{eqnarray}                        
R_{\eta^\prime}\equiv \frac{\Gamma(J/\psi\to\gamma\eta^\prime)}{\Gamma(\psi^\prime\to J/\psi\eta)}
=\bigg|\frac{{\bf q}(J/\psi\to\gamma\eta^\prime)}{{\bf q}(J/\psi\to \gamma\eta))}\bigg|^3
\bigg|\frac{m_{\eta^\prime}^2(\sqrt{2}\cos\theta_P+\sin\theta_P)}
{m_\eta^2(\cos\theta_P-\sqrt{2}\sin\theta_P)}\bigg|^2 R_\eta.
\label{R_eta'}
\end{eqnarray}
Taking $\theta_P\approx -20^\circ$ determined from the $\eta\to\gamma\gamma$ and 
$\eta^\prime\to\gamma\gamma$ rates, we predict
\begin{eqnarray}                          
R_{\eta^\prime}=0.044.
\label{R_eta'value}
\end{eqnarray}
The corresponding experimental value of $R_{\eta^\prime}$ is \cite{PDG}
\begin{eqnarray}                           
R_{\eta^\prime}\bigg|_{expt}=0.044\pm 0.010.
\label{R_eta'(expt)}
\end{eqnarray}
We see that this prediction is also in agreement with the experiment.

It has been shown in Ref.~\cite{KYF} that the contribution of the above mechanism 
to the isospin violating radiative decay $J/\psi\to\gamma+\pi^0$ is negligibly small.
There is another important mechanism giving the main contribution to $J/\psi\to\gamma+\pi^0$.
It is the $\rho^0$ meson dominance mechanism $J/\psi\to \rho^{0*}+\pi^0\to\gamma+\pi^0$. 
This has been studied in Ref.~\cite{FJ}, and the result is close to the experimental value
\footnote{The contribution of the $\rho^0$ meson dominance mechanism to $J/\psi\to\gamma\eta$
is negligibly small because the branching ratio $B(J/\psi\to\rho^0+\eta)$ is about two orders 
of magnitude smaller than $B(J/\psi\to\rho^0+\pi^0)$. }.

We would like
to mention that this approach is not suitable for
$\Upsilon\to\gamma\eta$ since the typical gluon momentum in this
process is $k\sim q_\eta/2\sim 2.4$ GeV at which perturbative QCD
works, while QCD multipole expansion does not. Studies of the processes
$\Upsilon\to\gamma\eta$, $\Upsilon\to\gamma\eta^\prime$ and
$\Upsilon\to\gamma f_2(1270)$
have been carried out in Ref. \cite{Ma}. Application of this
approach to $\psi^\prime\to\gamma\eta$ is more complicated since
both relativistic and coupled-channel corrections are important in
this process. Thus developing a relativistic coupled-channel is
desired.

QCDME can also be applied to study the direct
photon spectrum in $J/\psi\to\gamma+{\rm hadrons}~$ near $x\simeq
1$, where $x\equiv 2\omega_\gamma/M$ with $\omega_\gamma$ the
energy of the photon and $M$ the mass of the quarkonium.
Conventional study of the process $J/\psi(\Upsilon)\to\gamma+{\rm
hadrons}$~~are based on perturbative QCD calculation of
~$J/\psi(\Upsilon)\to\gamma+g+g$~ in the Born approximation
\cite{BDHC}. The direct photon spectrum is expressed as
\begin{eqnarray}                 
\frac{1}{\Gamma_{tot}}\frac{d\Gamma(J/\psi(\Upsilon)\to\gamma+{\rm hadrons})}{dx}.\nonumber
\end{eqnarray}
For $\Upsilon$, the obtained result is in good agreement with the
experiment \cite{CUSB,Photiadis,Field}, while for $J/\psi$ the
obtained distribution is too hard, i.e., in the range $x>0.8$, the
obtained distribution is much larger than the experimental values
\cite{Scharre}. In $J/\psi\to\gamma+{\rm hadrons}$, the typical
gluon momentum at $x\simeq 1$ is $k\simeq 770$ MeV, so that we can
apply QCDME to it for $x\ge 0.9$. The
calculation was done in Ref. \cite{CCKY}, and the obtained direct
photon spectrum for $x\ge 0.9$ is very close to the experimental values
\cite{CCKY}. For $x<0.9$, the typical gluon momentum is too large
for QCDME to work. A successful theory for the
whole range of $x$ is still expected.

\section{Summary and Outlook}

In this paper, we have reviewed the theory and applications of QCDME. 
We see from Secs.~III$\--$V that nonrelativistic QCDME theory gives many 
successful predictions for hadronic transition and some radiative decay rates in heavy quarkonium 
systems. Even the simple nonrelativistic single-channel theory can work well for many processes. 
Although the single-channel approach gives too small rates for $\Upsilon^\prime\to\Upsilon\pi\pi$, 
$\Upsilon^{\prime\prime}\to\Upsilon\pi\pi$, and $\Upsilon^{\prime\prime}\to\Upsilon^\prime\pi\pi$ 
(cf. TABLE~I), nonrelativistic coupled-channel theory improves the prediction
(cf. TABLE~IV). We summarize the above mentioned main successful predictions for the transition and 
decay rates in TABLE~V together with the corresponding experimental results for comparison.
\null\vspace{-0.3cm}
\begin{table}[h]
\caption{Summary of the predictions for transition and decay rates in the nonrelativistic QCD multipole expansion approach together
with the corresponding experimental results for comparison.}
\tabcolsep 5.5pt
\begin{tabular}{cccc}
\hline\hline
& Theoretical predictions&Experimental data&Places in the text\\
\hline
$\Gamma(\Upsilon^\prime\to\Upsilon~\pi\pi)$&13 keV&$12.0 \pm 1.8$ keV (PDG)&TABLE~IV\\
$\Gamma(\Upsilon^{\prime\prime}\to\Upsilon~\pi\pi)$&1.0 keV &$1.72\pm 0.35$ keV (PDG)&TABLE~IV\\
$\Gamma(\Upsilon^{\prime\prime}\to\Upsilon^\prime~\pi\pi)$&0.3 keV&$1.26\pm 0.40$ keV (PDG)&TABLE~IV\\
$\Gamma(\Upsilon^\prime\to\Upsilon~\eta)$&0.022 keV&$<0.086$ keV (PDG)&
Eqs.(\ref{Upsilon'Upsilon",eta})(\ref{eta-expt})\\
$\Gamma(\Upsilon^{\prime\prime}\to\Upsilon~\eta)$&0.011 keV&$<0.058$ keV (PDG)&
Eqs.(\ref{Upsilon'Upsilon",eta})(\ref{eta-expt})\\
$\displaystyle R^\prime\equiv\frac{\left|f^{111}_{2010}(b\bar{b})/m_b\right|^2
|\bm q(b\bar{b})|^3}
{\left|f^{111}_{2010}(c\bar{c})/m_c\right|^2|\bm q(c\bar{c})|^3}$&0.0025&$<0.0098$ (BES, PDG)&
Eqs.(\ref{R'R"})(\ref{R'R"expt})\\
$\displaystyle R^{\prime\prime}\equiv\frac{\left|f^{111}_{3010}(b\bar{b})/m_b\right|^2
|\bm q(b\bar{b})|^3}
{\left|f^{111}_{2010}(c\bar{c})/m_c\right|^2|\bm q(c\bar{c})|^3}$&0.0013&$<0.0065$ (BES, PDG)&
Eqs.(\ref{R'R"})(\ref{R'R"expt})\\
$\Gamma\big(\chi_b(2^3P_0)\to\chi_b(1^3P_0)~\pi\pi\big)$&0.4 keV&$\--$&TABLE~II\\
$\Gamma\big(\chi_b(2^3P_0)\to\chi_b(1^3P_2)~\pi\pi\big)$&$0.002\--0.02$ keV&$\--$&TABLE~II\\
$\Gamma\big(\chi_b(2^3P_1)\to\chi_b(1^3P_1)~\pi\pi\big)$&0.4 keV&$(0.83\pm0.22\pm 0.08\pm 0.19)$ keV
&TABLE~II\\
$\Gamma\big(\chi_b(2^3P_1)\to\chi_b(1^3P_2)~\pi\pi\big)$&$0.001\--0.01$ keV&$\--$&TABLE~II\\
$\Gamma\big(\chi_b(2^3P_2)\to\chi_b(1^3P_2)~\pi\pi\big)$&0.4 keV&$(0.83\pm0.22\pm 0.08\pm 0.19)$ keV
&TABLE~II\\
$\Gamma\big(\psi(3770)\to J/\psi~\pi^+\pi^-\big)$&$(32\--147)$ keV&$(80\pm32\pm21)$ keV (BES)&
TABLE~III,Eq.(\ref{BES3770-Gamma})\\
&&$(50.5\pm16.9)$ keV (CLEO-c)&Eq.(\ref{CLEO3770-Gamma})\\
$B(\psi^\prime\to h_c\pi^0)B(h_c\to\eta_c\gamma)$&$(1.9\--5.8)\times 10^{-4}$&
$(4\pm0.8\pm0.7)\times 10^{-4}$ (CLEO-c)&Eqs.(\ref{PQCDBB})(\ref{CLEO-cBB}) \\
$R_\eta\equiv\Gamma(J\psi\to\gamma\eta)/\Gamma(\psi^\prime\to J/\psi~\eta)$&0.012&$0.009\pm0.003$&
Eqs.(\ref{R_eta})(\ref{R_eta(expt)})\\
$R_{\eta^\prime}\equiv\Gamma(J\psi\to\gamma\eta^\prime)/\Gamma(\psi^\prime\to J/\psi~\eta)$&0.044&
$0.044\pm0.010$&
Eqs.(\ref{R_eta'})(\ref{R_eta'(expt)})\\
\hline\hline
\end{tabular}
\end{table}

In addition, the prediction for the $M_{\pi\pi}$ distribution in $\Upsilon^\prime\to\Upsilon\pi\pi$ 
from the nonrelativistic coupled-channel theory is in good agreement with the data (cf. FIG.~5). 

However, despite of the above success, there are experimental results which this simple 
nonrelativistic approach cannot explain. CLEO experiment shows a clear double-peak shape for the 
$M_{\pi\pi}$ distribution in $\Upsilon^{\prime\prime}\to\Upsilon~\pi\pi$ [cf. FIG.~6(a)]. 
Nonrelativistic coupled-channel correction is so small that it cannot account for this shape. Whether
this double-peak shape can be explained by the final state $\pi\pi$ interactions or it is caused by 
other physical effects is still not clear yet. Further investigation is needed.

Another problem is that, in the nonrelativistic single-channel approach, the $S$-wave 
to $S$-wave transitions rates are contributed only by the $C_1$ term in Eq.~(\ref{HofE1E1}) due to 
the orbital angular momentum selection rule, i.e., the obtained $\pi\pi$ angular correlation is 
isotropic in the laboratory frame.
However, experiments on $\psi^\prime\to J/\psi~\pi\pi$ \cite{BES01} shows a small angular dependence 
(a $0.2\%$ ingredient of $D$-wave) of the $\pi\pi$ angular correlation which cannot be explained by 
the nonrelativistic single-channel theory \footnote{Ref.~\cite{BES01} intended to use a theoretical 
formula given in Ref.~\cite{VZNS} to explain their data on the $\pi\pi$ angular correlarion by
making a direct comparison of that formula (given in the $\pi\pi$ rest frame in which $\psi^\prime$
is moving) with their partial wave analysis result from their data (done in the laboratory frame in 
which $\psi^\prime$ is at rest). However, such a comparison is actually inadequate. Since orbital 
angular momentum is not a Lorentz invariant quantity, partial wave decomposition of a transition 
amplitude is Lorentz frame dependent. Therefore it is not correct to directly compare the two partial 
wave decompositions obtained in different Lorentz frames. The correct way of doing it is to make a 
Lorentz transformation boosting that theoretical formula into the $\psi^\prime$ rest frame, and then 
make the comparison. It is easy to see that, after the Lorentz boost, the $D$-wave ingredient in the 
formula given in Ref.~\cite{VZNS} vanishes in the $\psi^\prime$ rest frame, i.e., the theoretical 
amplitude given in Ref.~\cite{VZNS} also leads to an isotropic $\pi\pi$ anglular correlation in the 
$\psi^\prime$ rest frame just as what Eq.~(\ref{HofE1E1}) does (with $C_2=0$). Thus an isotropic 
$\pi\pi$ angular correlation in the $\psi^\prime$ rest frame is a general consequence of all kinds 
of nonrelativistic single-channel approaches.}. Theoretically, the angular 
dependence of the transition rates may come from: (a) coupled-channel corrections [state-mixing leads 
to the $C_2$ term ($D$-wave) contributions], and (b) relativistic corrections [orbital angular momentum 
no longer conserves in the relativistic theory]. Actually, the sizes of the corrections 
(a) and (b) are of the same order of magnitude, and there is interference between them. 
Therefore to obtain a theoretical prediction for the $\pi\pi$ angular correlation, both (a) and (b) 
corrections should be taken into account. This means that a systematic relativistic coupled-channel 
theory of hadronic transitions is expected. So far there is still no such a theory due to the 
difficulty of dealing with the two-body bound-state equation in relativistic quantum mechanics. 
There have been various attempts to solve the relativistic two-body problem. Making effort
on developing a systematic relativistic coupled-channel theory of hadronic transition is
really important.

In a word the nonrelativistic theory of QCDME approach is not the end of the story.
Further development is needed.

\null\noindent
\begin{center}
{\bf Acknowledgment}
\end{center}

This work is supported by National Natural Science Foundation of China under Grant No. 90403017.

\end{document}